\documentclass[12pt]{article}
\usepackage{amssymb,amsmath,epsfig}

\begin{document}

\title{\bf Wave Properties of Plasma Surrounding
the Event Horizon of a Non-Rotating Black Hole}

\author{M. Sharif \thanks{msharif@math.pu.edu.pk} and G. Mustafa\\
Department of Mathematics,\\
University of the Punjab, Lahore 54590, Pakistan}
\date{}
\maketitle
\begin{abstract}
We have studied the wave properties of the cold and isothermal
plasma in the vicinity of the Schwarzschild black hole event
horizon. The Fourier analyzed perturbed 3+1 GRMHD equations are
taken on the basis of which the complex dispersion relations are
obtained for non-rotating, rotating non-magnetized and rotating
magnetized backgrounds. The propagation and attenuation vectors
along with the refractive index are obtained (shown in graphs) to
study the dispersive properties of the medium near the event
horizon. The results show that no information can be obtained from
the Schwarzschild magnetosphere. Further, the pressure ceases the
existence of normal dispersion of waves.
\end{abstract}


\section{Introduction}

Black holes are solutions of the Einstein field equations (EFEs) and
Schwarzschild solution is the simplest black hole solution. The most
physical black hole solution is the Kerr metric which is
axisymmetric and rotating. This reduces to the Schwarzschild black
hole when its angular momentum decreases to zero by energy
extraction \cite{1}. Around the Schwarzschild event horizon, the
powerful gravitational force pulls the magnetized plasma of the
surrounding space towards the event horizon in the form of an
accretion cloud. The accreting plasma creates a magnetic field. The
dynamical effects of accreting magnetospheric plasmas near a black
hole's event horizon enforce us to use the theory of general
relativistic magnetohydrodynamics (GRMHD) to study them.

The gravitomagetic waves and gravitational perturbations in the
black hole regime have always been of interest. Many people
\cite{3}-\cite{9} studied perturbations in the Schwarzschild
regime. To make the results of GR accessible for asytrophysicists,
Arnowitt, Deser and Misner \cite{10} developed a formulation (ADM
3+1 formalism) which splits the four-dimensional spacetime into
three-dimensional hypersurfaces labeled by time. This makes the
results of GR comparable with those of Newtonian physics. Several
authors \cite{11}-\cite{16} applied this formalism to judge
different aspects in GR.

Thorne and Macdonald \cite{17}-\cite{19} extended the formulation
to electromagnetic fields of black hole theory. Holcomb and Tajima
\cite{20}, Holcomb \cite{21} and Dettmann et al. \cite{22}
investigated the wave propagation in Friedmann universe. Buzzi et
al. \cite{23} discussed one-dimensional radial propagation of
transverse and longitudinal waves, in two component plasma, close
to the Schwarzschild event horizon. The stationary symmetric GRMHD
theory of black holes was developed by Zhang \cite{25}. He also
investigated the behavior of perturbations of cold plasma in the
ergosphere of Kerr black hole \cite{26}. Recently, Sharif and
Umber \cite{27}-\cite{31} have found some interesting wave
properties of cold and isothermal plasmas (with constant rest-mass
density) in the vicinity of Schwarzschild black hole event
horizon. They have evaluated real and complex wave numbers using
3+1 GRMHD equations.

In this paper, we shall consider variable rest-mass density to study
the wave properties of cold and isothermal plasmas in Schwarzschild
planar analogue by using complex wave vector components. The paper
has been organized as follows: Section \textbf{1} contains the
description of planar analogue with respective background,
perturbation and Fourier analysis assumptions. Sections \textbf{2},
\textbf{3} and \textbf{4} contain the dispersion relations obtained
for cold plasma living in non-rotating, rotating non-magnetized and
rotating magnetized backgrounds whereas sections \textbf{5},
\textbf{6} and \textbf{7} constitute the dispersion relations for
isothermal plasma living in the same backgrounds. All this has been
done for variable mass density and pressure. The results will be
discussed in the last section.

\section{Schwarzschild Planar Analogue and Relative Assumptions}

The Schwarzschild planar analogue can be described as \cite{26}
\begin{equation}\label{R}
ds^2=-\alpha^2(z)dt^2+dx^2+dy^2+dz^2.
\end{equation}
The directions $z$, $x$ and $y$ are analogous to Schwarzschild's
$r$, $\phi$ and $\theta$ respectively. In this planar analogue, we
assume the existence of cold and isothermal plasmas with the
respective equations of state \cite{26}
\begin{eqnarray}
\mu=\frac{\rho}{\rho_0}, \quad \mu=\frac{\rho+p}{\rho_0},
\end{eqnarray}
where $\rho_0,~\rho$ and $p$ denote the rest, moving mass
densities and pressure respectively.

In non-rotating background, the perturbed flow of fluid is only
along $z$-axis. The FIDO measured magnetic field and velocity are
given by
\begin{eqnarray}\label{g8}
\textbf{B}=B\textbf{e}_z,\quad \textbf{V}=u(z)\textbf{e}_z.
\end{eqnarray}
In rotating background, FIDO measured velocity and magnetic field
are two-dimensional. In planar analogue, these quantities can be
expressed in $xz$-plane by the following expressions
\begin{eqnarray}\label{g9}
\textbf{B}=B[\lambda(z)\textbf{e}_x+\textbf{e}_z],\quad
\textbf{V}=V(z)\textbf{e}_x+u(z)\textbf{e}_z.
\end{eqnarray}
Here $\lambda$, $u$ and $V$ are related to each other by the
following equation \cite{27}
\begin{equation}\label{r}
V=\frac{V^F}{\alpha}+\lambda u,
\end{equation}
where $V^F$ is an integration constant.

We assume linear (first order) perturbations in flow variables
(mass density $\rho$, pressure $p$, velocity $\textbf{V}$ and
magnetic field $\textbf{B}$) of the fluid.
\begin{eqnarray}\label{a5}
&&\rho=\rho^0+\delta\rho=\rho^0+\rho\widetilde{\rho},\quad
p=p^0+\delta p=p^0+p\widetilde{p},\nonumber\\
&&\textbf{V}=\textbf{V}^0+\delta\textbf{V}=\textbf{V}^0+\textbf{v},\quad
\textbf{B}=\textbf{B}^0+\delta\textbf{B}=\textbf{B}^0+B\textbf{b},
\end{eqnarray}
where $\rho^0,~p,~\textbf{V}^0$ and $\textbf{B}^0$ are unperturbed
quantities, $\delta\rho,~\delta p,\delta\textbf{V}$ and
$\delta\textbf{B}$ represent perturbed quantities. The
dimensionless perturbed quantities
$\widetilde{\rho},~\widetilde{p},~\textbf{v}$ and $\textbf{b}$ can
be written as follows
\begin{eqnarray}\label{g10}
&&\tilde{\rho}=\tilde{\rho}(t,z),\quad \tilde{p}=\tilde{p}(t,z),\nonumber\\
&&\textbf{v}=\delta\textbf{V}=v_x(t,z)\textbf{e}_x
+v_z(t,z)\textbf{e}_z,\nonumber\\
&&\textbf{b}=\frac{\delta\textbf{B}}{B}=b_x(t,z)\textbf{e}_x
+b_z(t,z)\textbf{e}_z.
\end{eqnarray}
For Fourier analysis, we assume the harmonic space and time
dependence of perturbations
\begin{eqnarray}\label{j12}
\widetilde{\rho}(t,z)=c_1e^{-\iota(\omega t-kz)},&\quad&
\widetilde{p}(t,z)=c_2e^{-\iota(\omega t-kz)},\nonumber\\
v_z(t,z)=c_3e^{-\iota(\omega t-kz)},&\quad&
v_x(t,z)=c_4e^{-\iota(\omega
t-kz)},\nonumber\\
b_z(t,z)=c_5e^{-\iota(\omega t-kz)},&\quad&
b_x(t,z)=c_6e^{-\iota(\omega t-kz)}.
\end{eqnarray}

\section{3+1 Perfect GRMHD Equations in Planar Analogue}

For the Schwarzschild planar analogue, the perfect GRMHD equations
can be written as \cite{27},\cite{ut}
\begin{eqnarray}\setcounter{equation}{1}\label{z1}
&&\frac{\partial\textbf{B}}{\partial t}=\nabla \times(\alpha
\textbf{V}\times\textbf{B}),\\
\label{z2} &&\nabla.\textbf{B}=0,\\
\label{z3} &&\frac{\partial\rho}{\partial
t}+{(\alpha\textbf{V}).\nabla}\rho+\rho\gamma^2
\textbf{V}.\frac{\partial\textbf{V}}{\partial
t}+\rho\gamma^2\textbf{V}.(\alpha\textbf{V}.\nabla)\textbf{V}
\nonumber\\
&&+\rho{\nabla.(\alpha\textbf{V})}=0,\\
\label{z4}&&\{(\rho\gamma^2+\frac{\textbf{B}^2}{4\pi})\delta_{ij}
+\rho\gamma^4V_iV_j
-\frac{1}{4\pi}B_iB_j\}(\frac{1}{\alpha}\frac{\partial}{\partial
t}+\textbf{V}.\nabla)V^j\nonumber\\
&&-(\frac{\textbf{B}^2}{4\pi}\delta_{ij}-\frac{1}{4\pi}B_iB_j)
V^j,_kV^k+\rho_0\gamma^2V_i\{\frac{1}{\alpha}\frac{\partial
\mu}{\partial
t}+(\textbf{V}.\nabla)\mu\}\nonumber\\
&&=-\rho\gamma^2a_i-p,_i+
\frac{1}{4\pi}(\textbf{V}\times\textbf{B})_i\nabla.(\textbf{V}\times\textbf{B})
-\frac{1}{8\pi\alpha^2}(\alpha\textbf{B})^2,_i\nonumber\\
&&+\frac{1}{4\pi\alpha}(\alpha B_i),_jB^j-\frac{1}{4\pi\alpha}
[\textbf{B}\times\{\textbf{V}\times(\nabla\times(\alpha\textbf{V}
\times\textbf{B}))\}]_i,\\
\label{z5} &&\gamma^2 (\frac{1}{\alpha} \frac{\partial}{\partial
t}+\textbf{V}.\nabla)(\mu\rho_0)-\frac{1}{\alpha}\frac{\partial
p}{\partial t}
+2\rho_0\mu\gamma^4\textbf{V}.(\frac{1}{\alpha}\frac{\partial}{\partial
t}+\textbf{V}.\nabla)\textbf{V}\nonumber
\end{eqnarray}
\begin{eqnarray}
&&+2\rho_0\mu\gamma^2(\textbf{V}.\textbf{a})
+\rho_0\mu\gamma^2(\nabla.\textbf{V})
+\frac{1}{4\pi\alpha}[(\textbf{V}\times\textbf{B}).
(\nabla\times(\alpha\textbf{B}))\nonumber\\&&+(\textbf{V}\times\textbf{B})
.\frac{\partial}{\partial
t}(\textbf{V}\times\textbf{B})].
\end{eqnarray}
The cold plasma model can be described by
Eqs.(\ref{z1})-(\ref{z4}) whereas isothermal plasma needs
Eq.(\ref{z5}) for a better interpretation.

In a stationary symmetric background, the observers move along
symmetric directions and thus do not see any change in flow around
themselves. For such magnetized fluids, Phinney proposed the stream
functions $h$, $l$ and $e$ \cite{25}, i.e.,
\begin{eqnarray}\label{v1}
h=\frac{-4\pi u}{B},\quad l=\gamma V-\alpha \lambda B/h,\quad
e=\gamma\alpha,
\end{eqnarray}
where the rest-mass conservation law in three dimensions is
$\alpha\rho_0\gamma u=-1$ with specific enthalpy $\mu=1$ and the
constant $V^F=0.$

\section{Non-Rotating Background with Cold Plasma}

The Fourier analyzed perturbed GRMHD equations in non-rotating
background (Eqs.(3.15)-(3.18) of \cite{27}) are

\begin{eqnarray}\setcounter{equation}{1}\label{c9}
&&-\frac{\iota\omega}{\alpha}c_5=0,\\
\label{d1}&&\iota kc_5=0,\\
\label{d2}&&c_1\left(\frac{-\iota\omega}{\alpha}+\iota ku\right)
+c_3\left\{(1+\gamma^2u^2)\iota
k-(1+\gamma^2u^2)(1-2\gamma^2u^2)\frac{u'}{u} \right.\nonumber
\\&&\left.-\frac{\iota\omega}{\alpha}\gamma^2u\right\}=0,\\
\label{d3}&&c_1\gamma^2\{a_z+uu'(1+\gamma^2u^2)\}+c_3\left[\gamma^2(1+\gamma^2u^2)
\left(\frac{-\iota\omega}{\alpha}+ \iota uk\right)\right.\nonumber\\
&&\left.+\gamma^2\{u'(1+\gamma^2u^2)(1+4\gamma^2u^2)+2\gamma^2ua_z\}\right]=0.
\end{eqnarray}
Equations (\ref{c9}) and (\ref{d1}) show that $c_5=0$ which means
that there are no perturbations in magnetic field. Thus the
non-magnetized as well as magnetized backgrounds admit the same
perturbed Fourier analyzed GRMHD equations.

\subsection{Numerical Solutions}

In order to find the numerical solutions, we assume the time lapse
$\alpha=\frac{1}{10}\textmd{tanh(10z)}$. For stationary flow, we
take $\alpha\gamma=1$ which implies that $\gamma=1/\alpha$. For
the inflow of fluid into the black hole event horizon we take
$u=-{\sqrt{1-\alpha^2}}$. Using these assumptions, the mass
conservation law in three dimensions gives $\rho=-\frac{1}{u}.$
These quantities satisfy the GRMHD equations for the region
$0.5\leq z\leq10$. For these assumptions, $e=1$.

The dispersion relation from Eqs.(\ref{d2}) and (\ref{d3}) can be
obtained (using \emph{Mathematica}) by equating determinant of the
coefficients of constants $c_1$ and $c_3$ to zero \cite{35}. This
determinant leads to a complex dispersion relation of the form
\begin{equation}\label{v5}
A_1(z,\omega)k^2+A_2(z,\omega)k+A_3(z,\omega)+\iota\{A_4(z,\omega)k
+A_5(z,\omega)\}=0
\end{equation}
which gives two complex values of $k$. Since $k$ is the
$z$-component of the wave vector, it gives quantities
corresponding to $z$-direction. The real and imaginary parts of
$k$ yield the propagation and attenuation vectors respectively.
The propagation vector gives refractive index on the basis of
which the mode of dispersion can be found. The sinusoidal
expression then takes the form $e^{-\iota(\omega t-k_1z-\iota
k_2z)}=e^{-\iota(\omega t-k_1z)-k_2z}$, where $k_1=\textmd{Re}(k)$
and $k_2=\textmd{Im}(k).$ The two values of $k$ are shown in
Figure \textbf{1} and \textbf{2}.

\begin{figure}
\center \epsfig{file=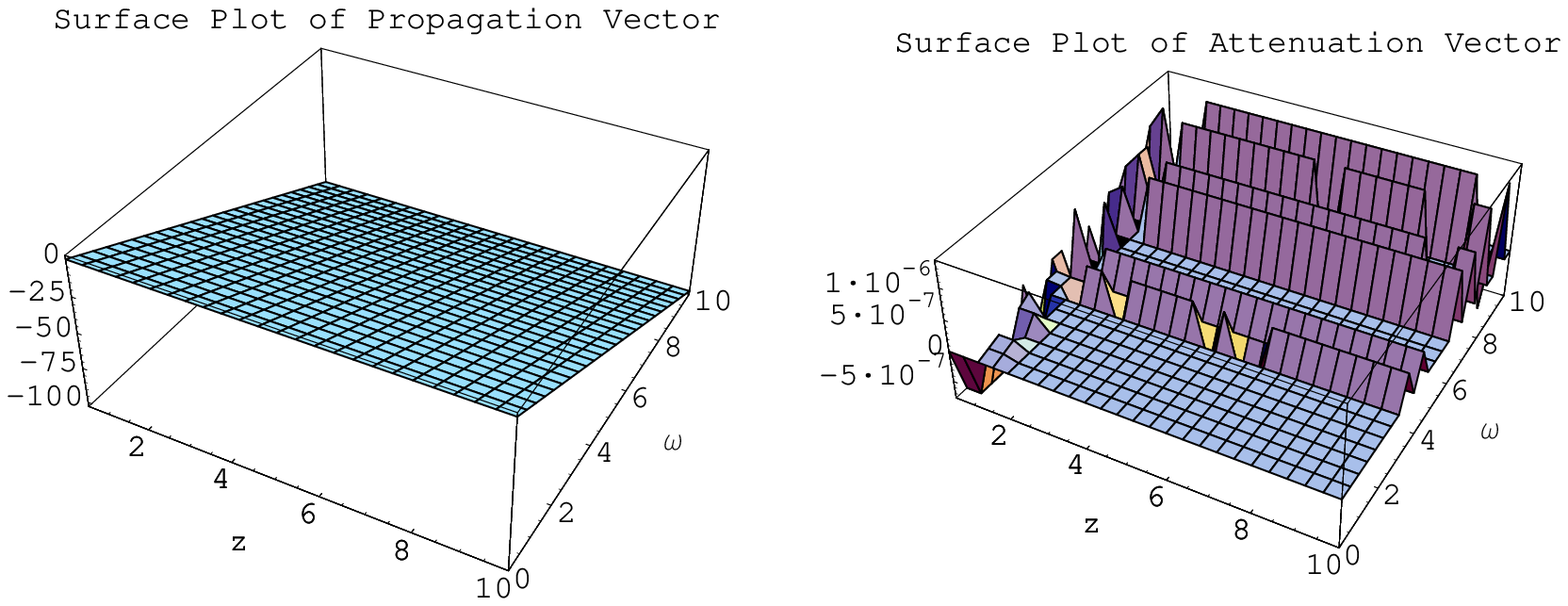,width=0.7\linewidth} \center
\epsfig{file=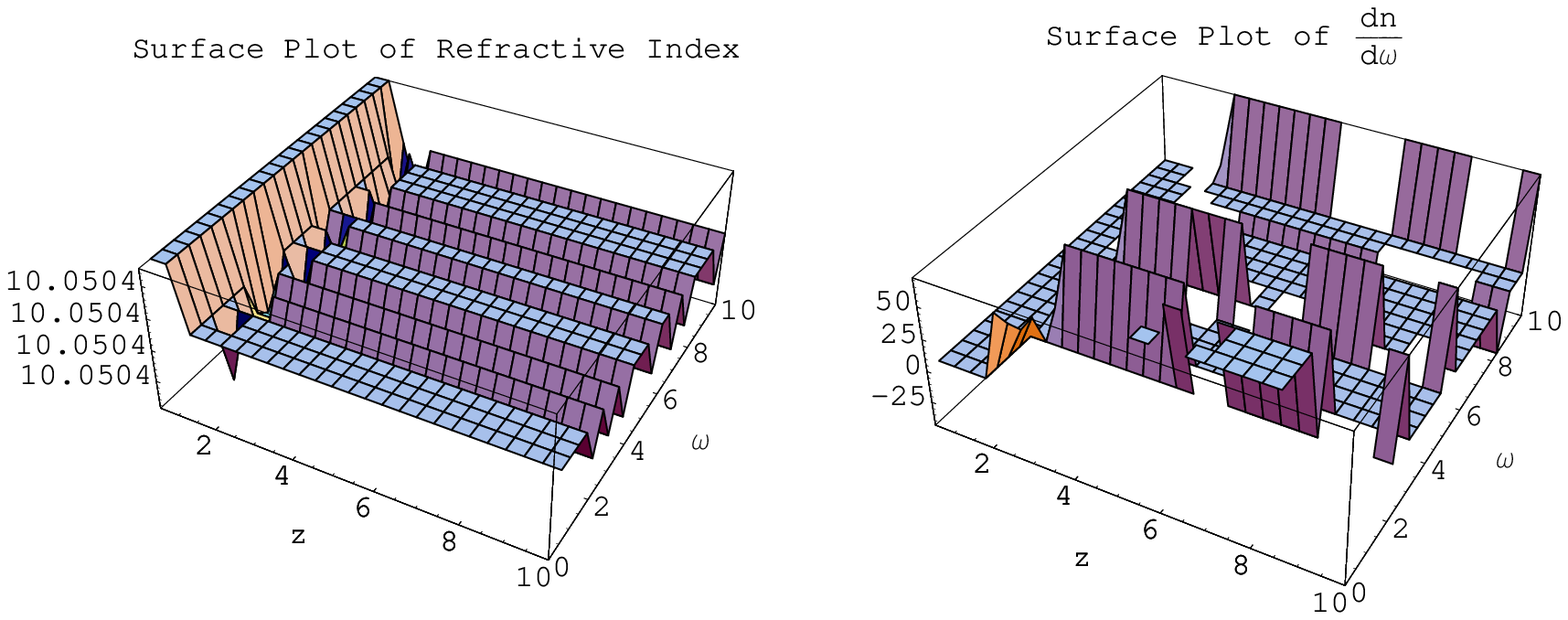,width=0.7\linewidth} \caption {The waves are
directed towards the event horizon. Dispersion is normal in most
of the region}
\end{figure}

Figure \textbf{1} indicates that the propagation vector decreases
with the increase in $z$ and $\omega$. It means that the waves
propagate slowly as they move away from the event horizon and as
their angular frequency increase. The attenuation vector admits
positive and negative values randomly which shows random damping
and growth of waves. The refractive index is greater than one and
its variation with respect to angular frequency is positive in the
region $1.9\leq z\leq10, ~1\leq\omega \leq10$ which gives that the
region admits normal dispersion of waves \cite{37}.
\begin{figure}
\center \epsfig{file=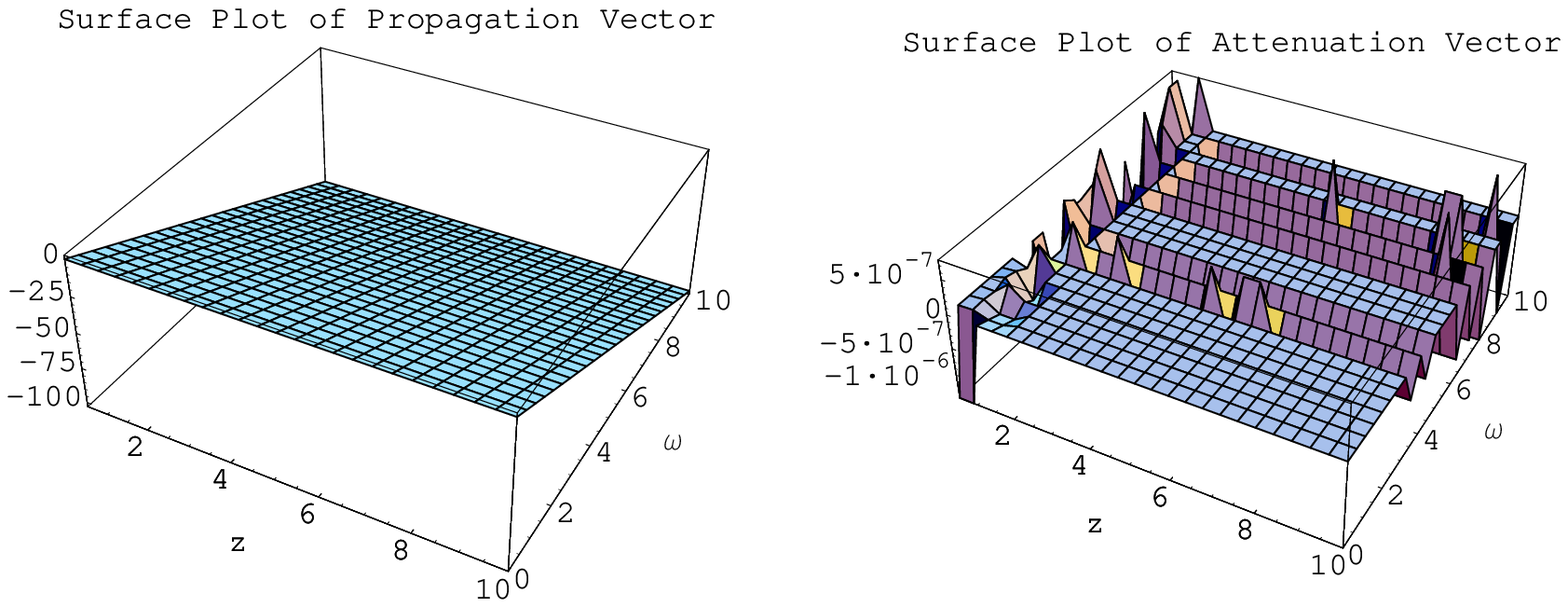,width=0.7\linewidth} \center
\epsfig{file=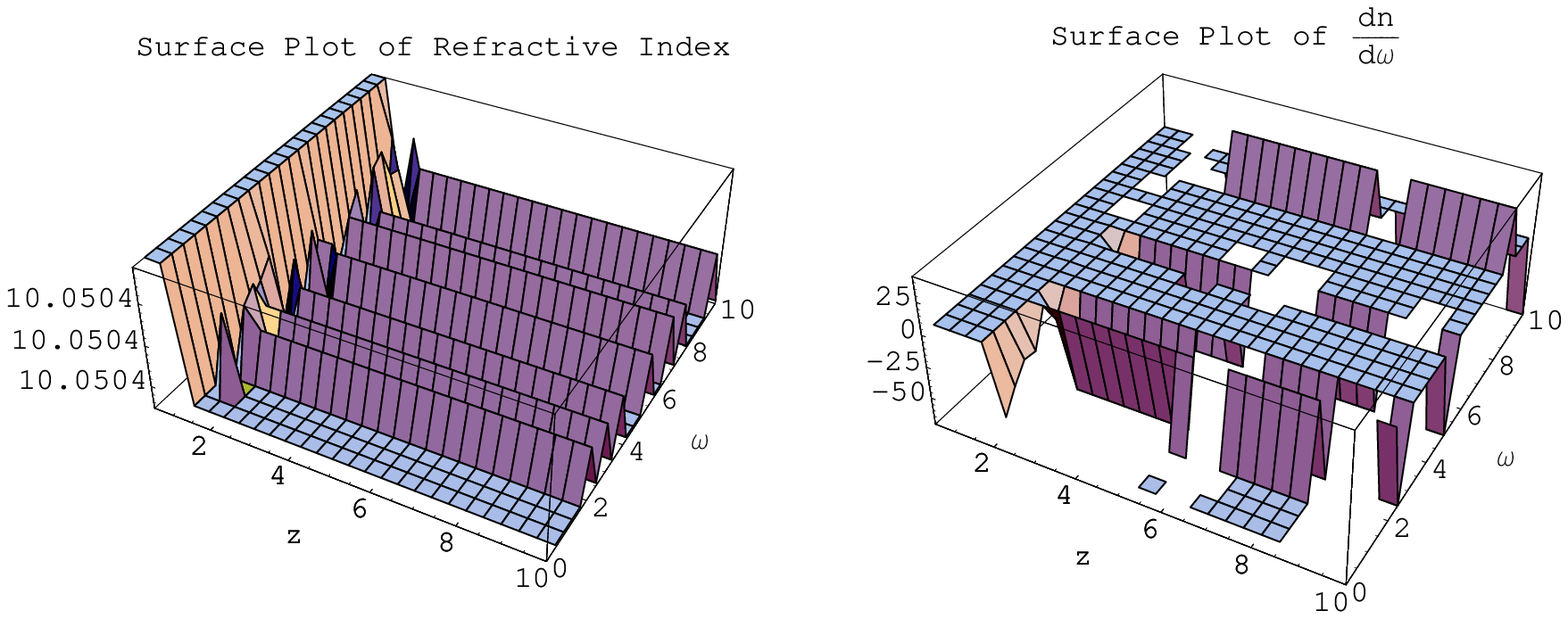,width=0.7\linewidth} \caption{The waves move
towards the event horizon. Most of the region admits anomalous
dispersion of waves}
\end{figure}
Figure \textbf{2} shows that the attenuation vector takes positive
values at random points. The propagation vector decreases with the
increase in angular frequency. The attenuation vector takes random
values which means that the waves damp and grow randomly with the
increase in angular frequency. The refractive index is greater
than one but $\frac{dn}{d\omega}<0$ in most of the region giving
anomalous dispersion of waves \cite{37}.

In Figures \textbf{1} and \textbf{2}, the propagation vector takes
negative values which shows that the waves move towards the black
hole event horizon. In addition, a small region near the event
horizon admits decrease in refractive index with the increase in
$z$.

\section{Rotating, Non-magnetized Background with Cold Plasma}

For the rotating non-magnetized background, the Fourier analyzed
perturbed perfect GRMHD equations are given by Eqs.(4.13)-(4.15)
of \cite{27}.
\begin{eqnarray}\setcounter{equation}{1}
\label{d7} &&c_1\left(\frac{-\iota\omega}{\alpha}+\iota uk\right)+
c_3\left[\frac{-\iota\omega}{\alpha}\gamma^2u+(1+\gamma^2u^2)\iota
k-(1-2\gamma^2u^2)\right.\nonumber\\
&&\left.\times(1+\gamma^2u^2)\frac{u'}{u}
+2\gamma^4u^2VV'\right]+c_4\gamma^2\left[\left(\frac{-\iota\omega}{\alpha}+\iota
ku\right)V+\gamma^2u\right.\nonumber\\
&&\left.\times\{(1+2\gamma^2V^2)V'+2\gamma^2uVu'\}\right]=0,\\
\label{d8} &&c_1\gamma^2u\{(1+\gamma^2V^2)V'+\gamma^2uVu'\}
+c_3\gamma^2\left[\left(\frac{-\iota\omega}{\alpha}+\iota
uk\right)\gamma^2uV\right.\nonumber\\
&&\left.+\{(1+2\gamma^2u^2)(1+2\gamma^2V^2)-\gamma^2V^2\}V'
+2\gamma^2(1+2\gamma^2u^2)uVu'\right]\nonumber\\
&&+c_4\left[\left(\frac{-\iota\omega}{\alpha}+\iota
ku\right)\gamma^2(1+\gamma^2V^2)+\gamma^4u\{(1+4\gamma^2V^2)uu'
\right.\nonumber\\
&&\left.+4VV'(1+\gamma^2V^2)\}\right]=0,\\
\label{d9}&&c_1\gamma^2\{a_z+(1+\gamma^2u^2)uu'+\gamma^2u^2VV'\}
+c_3\left[\gamma^2(1+\gamma^2u^2)
\left(\frac{-\iota\omega}{\alpha}\right.\right.\nonumber
\\&&\left.\left.+\iota
uk\right)+\gamma^2[u'(1+\gamma^2u^2)(1+4\gamma^2u^2)
+2u\gamma^2\{a_z+(1+2\gamma^2u^2)VV'\}]\right]
\nonumber\\
&&+c_4\gamma^4\left[\left(\frac{-\iota\omega}{\alpha}+\iota
uk\right)uV+u^2V'(1+4\gamma^2V^2)\right.\nonumber\\
&&\left.+2V(a_z+(1+2\gamma^2u^2)uu')\right]=0.
\end{eqnarray}

\subsection{Numerical Solutions}

We assume the same time lapse which we have assumed for
non-rotating background. The stationary flow assumption
$\alpha\gamma=1$ with $V=u$ leads to
$u=-\sqrt{\frac{1-\alpha^2}{2}}.$ Thus the mass conservation law
in three dimensions gives $\rho=-\frac{1}{u}.$ These values
satisfy the perfect GRMHD equations for the range $0.75\leq
z\leq10$. In rotating non-magnetized background, the constant
values of $u=V=-0.703562$ make the flow constants $l=-7.03562$ and
$e=1$. Consequently, we obtain the following form of complex
dispersion relation cubic in $k$.
\begin{eqnarray}
&&A_1(z)k^2+A_2(z,\omega)k+A_3(z,\omega)+\iota\{A_4(z)k^3+A_5(z,\omega)k^2
\nonumber\\
&&+A_6(z,\omega)k +A_7(z,\omega)\}=0
\end{eqnarray}
This dispersion relation gives three complex values of $k$ with
corresponding graphs shown in Figures \textbf{3}-\textbf{5}.

\begin{figure} \center
\epsfig{file=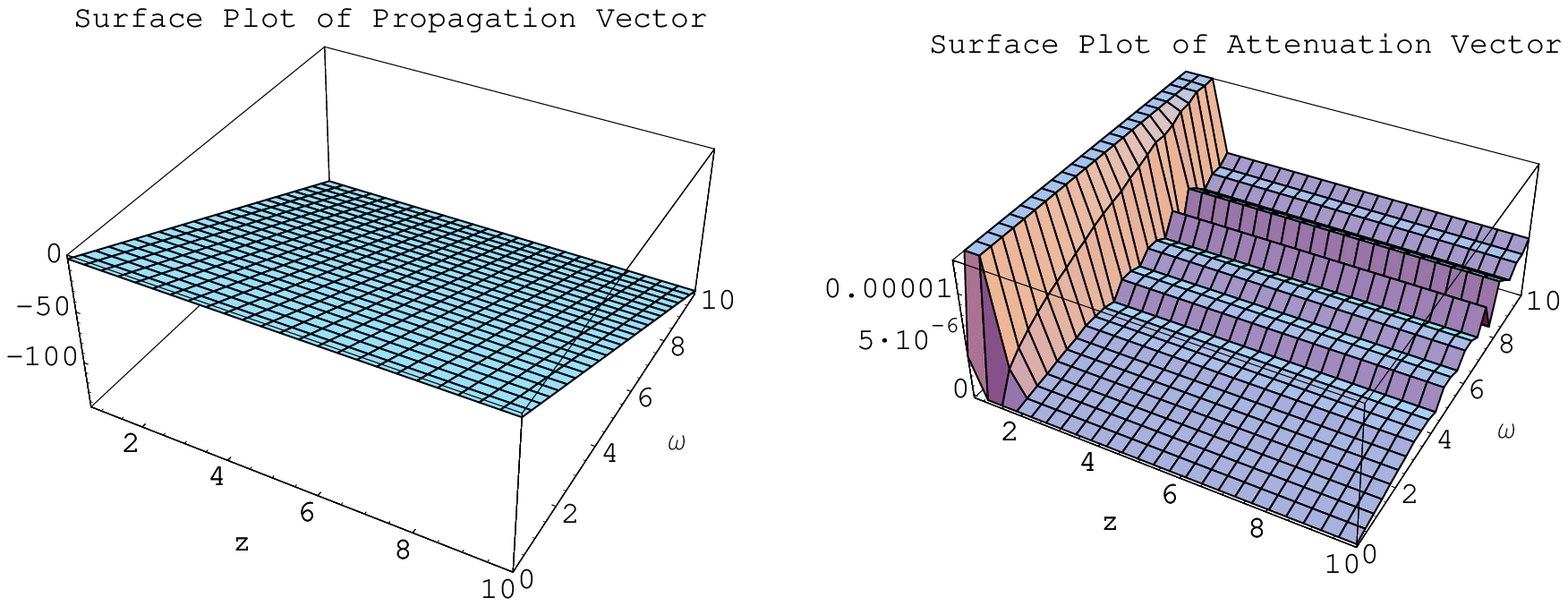,width=0.7\linewidth} \center
\epsfig{file=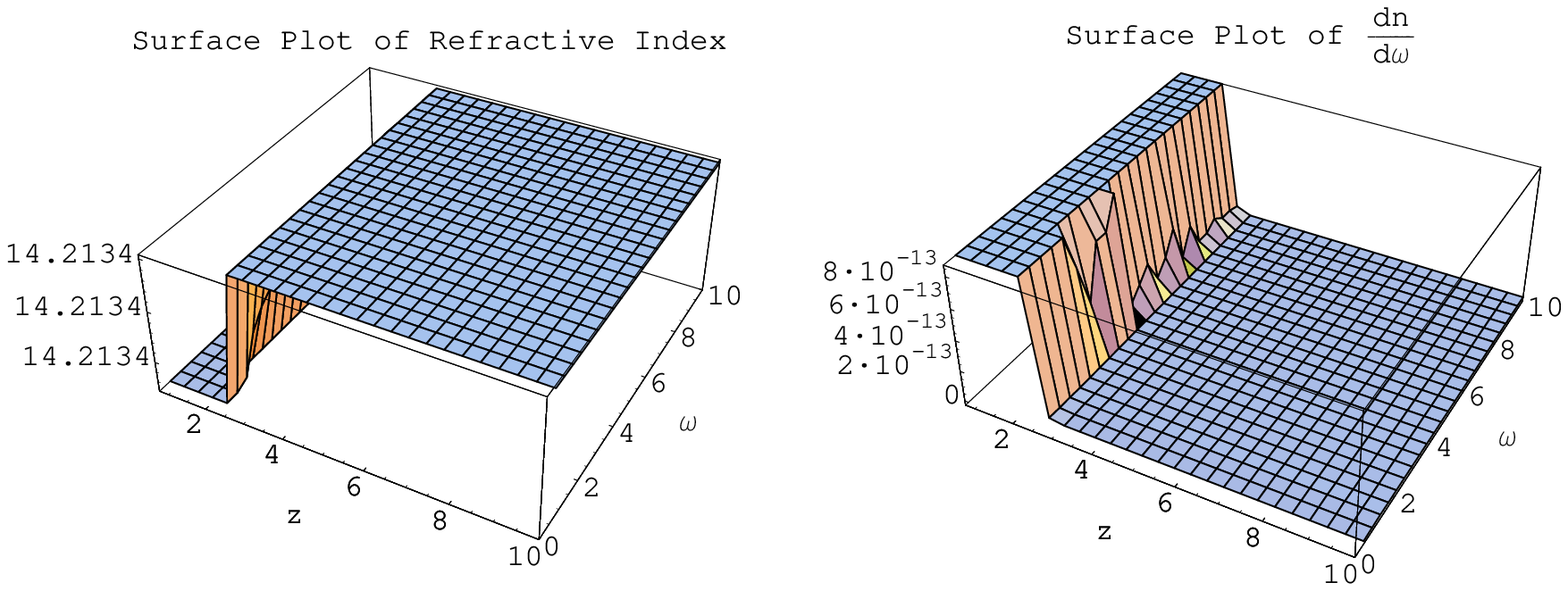,width=0.7\linewidth} \caption{The waves move
towards the event horizon. Most of of the region shows normal
dispersion of waves}
\end{figure}
In Figure \textbf{3}, the propagation vector decreases with the
increase in $z$. The attenuation vector decreases with the
increase in $z$ in a small region near the event horizon. Thus, in
the small region near the event horizon, the waves damp as they
move towards the event horizon. The refractive index is greater
than one and $\frac{dn}{d\omega}>0$ except for the region $2.8\leq
z\leq10, ~1\leq\omega \leq2.1$ which shows normal dispersion of
waves.
\begin{figure}
\center \epsfig{file=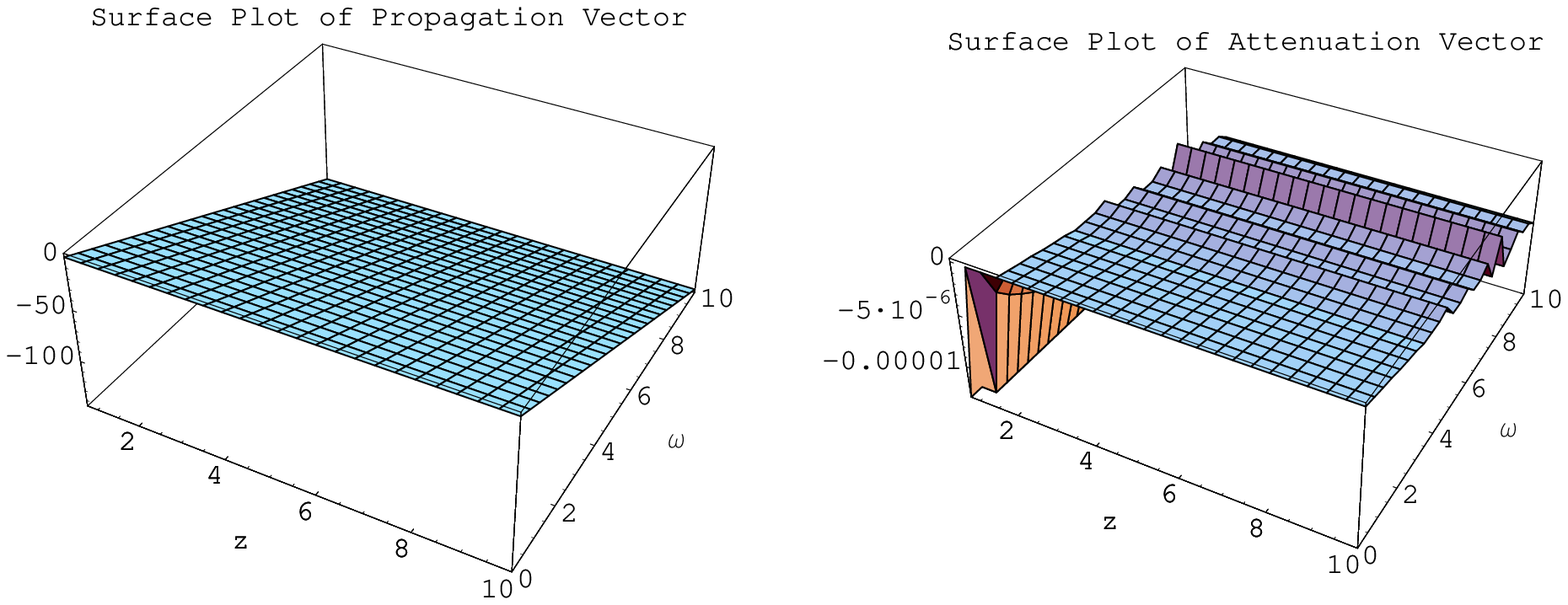,width=0.7\linewidth} \center
\epsfig{file=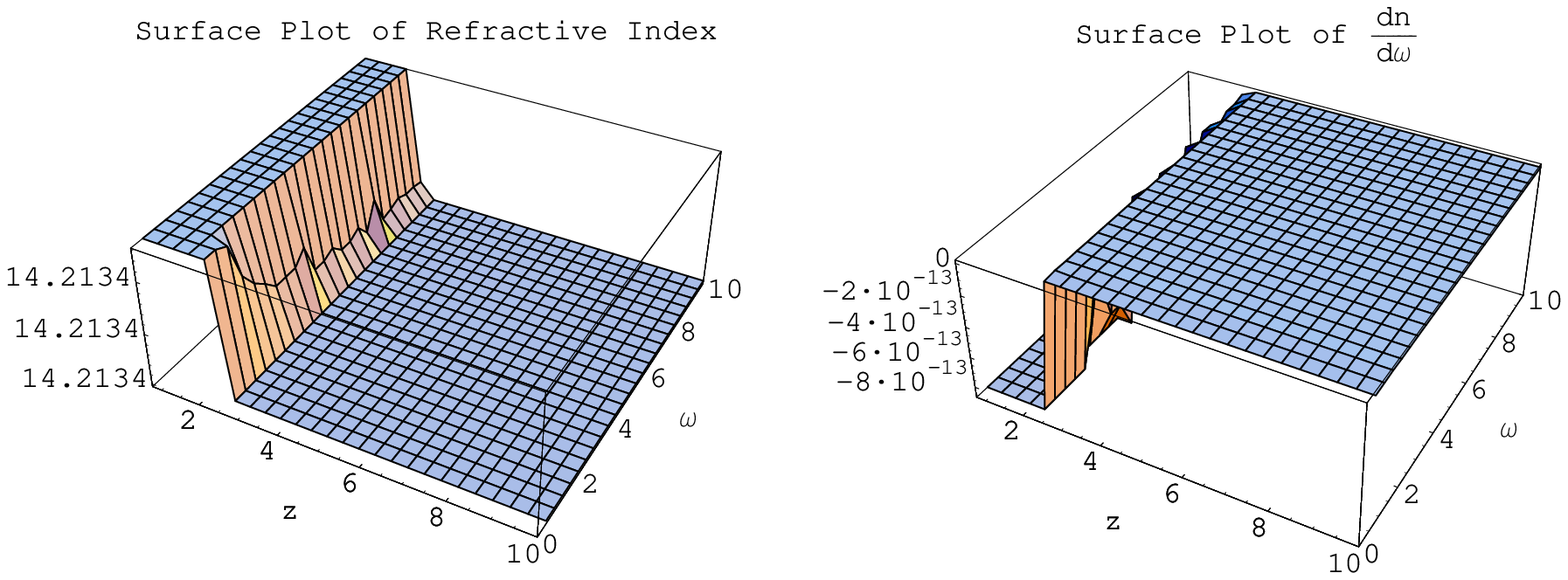,width=0.7\linewidth} \caption{The waves move
towards the event horizon. The region shows anomalous dispersion}
\end{figure}
Figure \textbf{4} shows that the attenuation vector takes random
negative values with increase in angular frequency. In a small
region near the event horizon, the attenuation vector increases
with increase in $z$. This indicates that the waves grow as they
move towards the event horizon. The refractive index is greater
than one and its variation with respect to angular frequency is
negative which shows that the whole region admits anomalous
dispersion of waves.
\begin{figure}
\center \epsfig{file=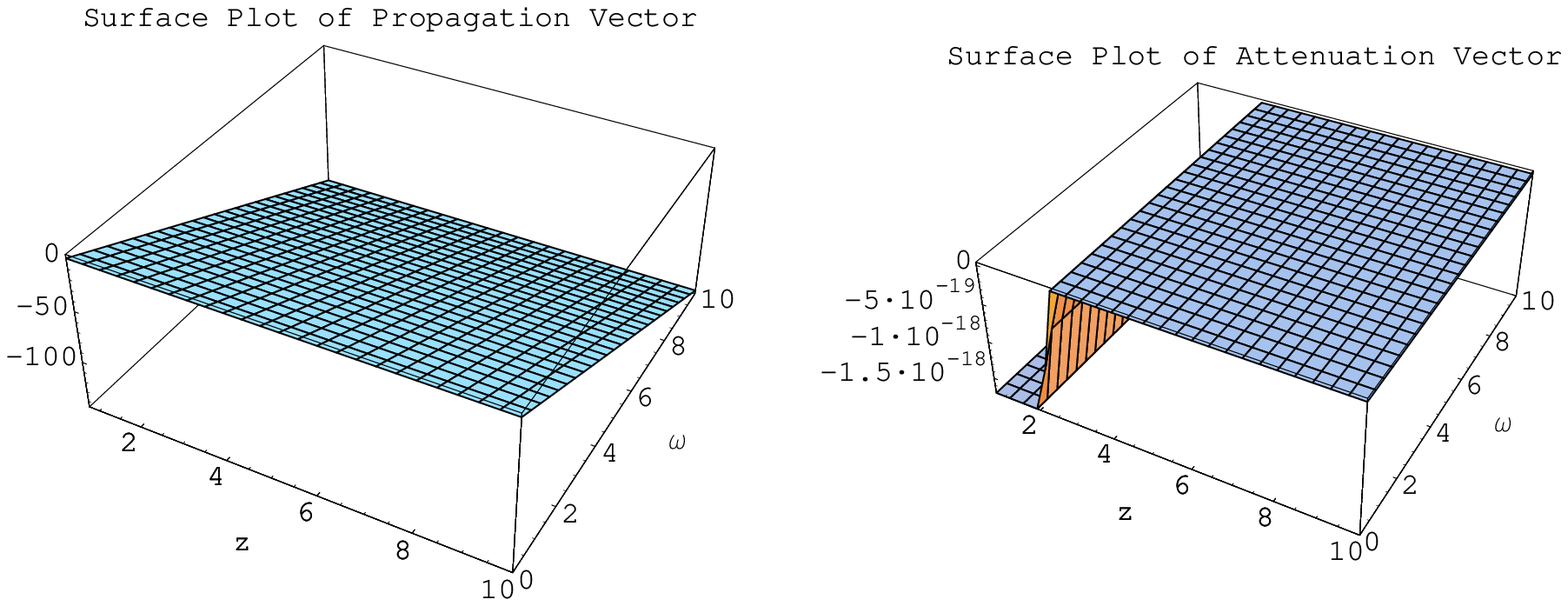,width=0.7\linewidth} \center
\epsfig{file=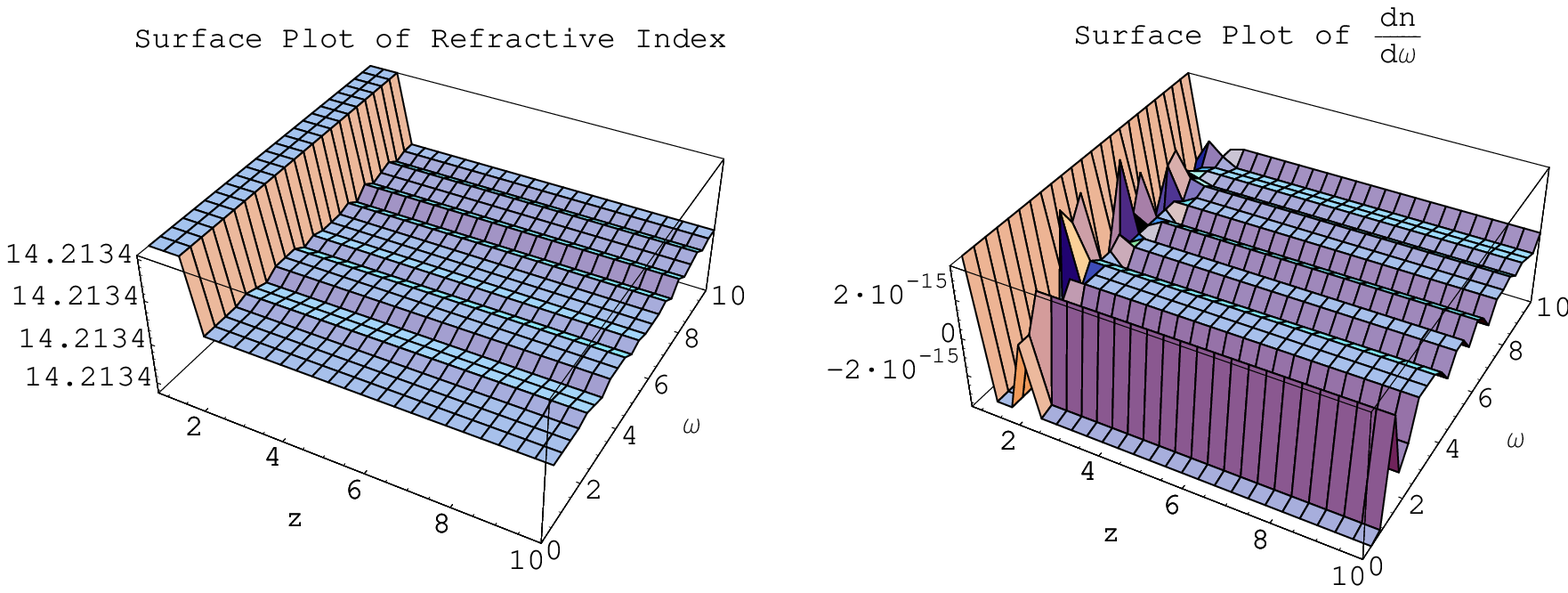,width=0.7\linewidth} \caption{The waves are
directed towards the the event horizon. A small region near the
event horizon admits normal dispersion of waves}
\end{figure}
Figure \textbf{5} indicates that the attenuation vector takes
negative values. Near the event horizon, the attenuation vector
decreases with the increase in $z$. It means that waves grow near
the event horizon. The refractive index is greater than one and its
variation with respect to angular frequency is greater than zero in
the region $0.75\leq z\leq1.1, ~1.5 \times 10^{-5}\leq\omega \leq
10$ which admits normal dispersion of waves. In rest of the region,
dispersion is anomalous and normal randomly due to variation in the
refractive index with respect to angular frequency which attain
positive and negative values randomly respectively.

In Figures \textbf{3}-\textbf{5}, the propagation is negative in
the whole region. This quantity decreases with the increase in
angular frequency. In Figures \textbf{4} and \textbf{5}, a small
region near the event horizon shows that the refractive index
decreases with the increase in $z$.

\section{Rotating Magnetized Background with Cold Plasma}

In this background, the Fourier analyzed form of the perturbed
GRMHD equations (Eqs.(5.15)-(5.20) of \cite{27}) is given as
follows
\begin{eqnarray}\setcounter{equation}{1}\label{g11}
&&c_4(\alpha'+\iota k\alpha)-c_3\{(\alpha\lambda)'+\iota
k\alpha\lambda\}+c_5(\alpha V)'-c_6\{(\alpha
u)'-\iota\omega\nonumber\\&&+\iota ku\alpha\}=0,\\
\label{g12}&&c_5(\frac{-\iota\omega}{\alpha}+\iota ku)=0,\\
\label{g13}&&c_5\iota k=0,\\
\label{g14}&&c_1(\frac{-\iota\omega}{\alpha}+\iota uk)+
c_3[\frac{-\iota\omega}{\alpha}\gamma^2u+(1+\gamma^2u^2)\iota
k-(1-2\gamma^2u^2)\nonumber\\&&\times(1+\gamma^2u^2)\frac{u'}{u}
+2\gamma^4u^2VV']+c_4\gamma^2[\frac{-\iota\omega}{\alpha}V+ \iota
kuV\nonumber\\&&+u\{(1+2\gamma^2V^2)V'+2\gamma^2uVu'\}]=0,\\
\label{g15}&&c_1\gamma^2\rho u\{(1+\gamma^2V^2)V'
+\gamma^2uVu'\}-\frac{B^2}{4\pi}c_6\{(1-u^2)\iota
k+\frac{\alpha'}{\alpha}(1-u^2)\nonumber\\
&&- uu'\}+c_3[-(\rho\gamma^4uV-\frac{\lambda
B^2}{4\pi})\frac{\iota\omega}{\alpha}+(\rho\gamma^4uV+\frac{\lambda
B^2}{4\pi})\iota ku\nonumber\\
&&+\rho\gamma^2\{(1+2\gamma^2u^2)(1+2\gamma^2V^2)
-\gamma^2V^2\}V'+2\rho\gamma^4(1+2\gamma^2u^2)u V
u'\nonumber\\
&&+\frac{B^2u}{4\pi\alpha}(\alpha\lambda)']
+c_4[-\{\rho\gamma^2(1+\gamma^2V^2)+\frac{B^2}{4\pi}\}
\frac{\iota\omega}{\alpha}+\{\rho\gamma^2(1+\gamma^2V^2)\nonumber
\end{eqnarray}
\begin{eqnarray}
&&- \frac{B^2}{4\pi}\}\iota ku+\rho\gamma^4u\{(1+4\gamma^2V^2)uu'
+4VV'(1+\gamma^2V^2)\}\nonumber\\
&&-\frac{B^2u\alpha'}{4\pi \alpha }]=0,\\
\label{g16}&&c_1\rho\gamma^2\{a_z+(1+\gamma^2u^2)uu'
+\gamma^2u^2VV'\}+c_3[-\{\rho\gamma^2(1+\gamma^2u^2)\nonumber\\
&&+\frac{\lambda^2B^2}{4\pi}\}\frac{\iota\omega}{\alpha}
+\{\rho\gamma^2(1+\gamma^2u^2)-\frac{\lambda^2B^2}{4\pi}\}\iota
ku+\{\rho\gamma^2\{u'(1+\gamma^2u^2)\nonumber\\&&\times(1+4\gamma^2u^2)
+2u\gamma^2(a_z+(1+2\gamma^2u^2)VV')\}-\frac{\lambda
B^2u}{4\pi\alpha}(\alpha\lambda)'\}]\nonumber\\
&&+c_4[-(\rho\gamma^4uV-\frac{\lambda
B^2}{4\pi})\frac{\iota\omega}{\alpha}+(\rho\gamma^4uV+\frac{\lambda
B^2}{4\pi})\iota ku\nonumber\\
&&+\{\rho\gamma^4\{u^2V'(1+4\gamma^2V^2)
+2V(a_z+(1+2\gamma^2u^2)uu')\}+\frac{\lambda
B^2\alpha'u}{4\pi\alpha}\}]\nonumber\\&&+\frac{B^2}{4\pi}c_6[\lambda(1-u^2)\iota
k+\lambda\frac{\alpha'}{\alpha}(1-u^2)-\lambda
uu'+\frac{(\alpha\lambda)'}{\alpha}]=0.
\end{eqnarray}
Equations (\ref{g12}) and (\ref{g13}) give $c_5=0$ which shows
that there are no perturbations in $z$-component of magnetic
field.

\subsection{Numerical Solutions}

We assume the same values of time lapse, density, $x$ and
$z$-component of velocity which we have mentioned in the previous
section. Substituting $u=V$ in Eq.(\ref{r}) with the assumption
that $V^F=0$, it gives that $\lambda=1$. Also we take
$B=\sqrt{\frac{176}{7}}$. These values satisfy the GRMHD equations
for the region $1\leq z\leq10$. For this region, the numerical
values of the flow constants (due to constant $u=V=-0.703562$, $B$
and $\lambda$) are $e=1$, $l=-6.83562$ and $h=-2.50663$.

\begin{figure}
\center \epsfig{file=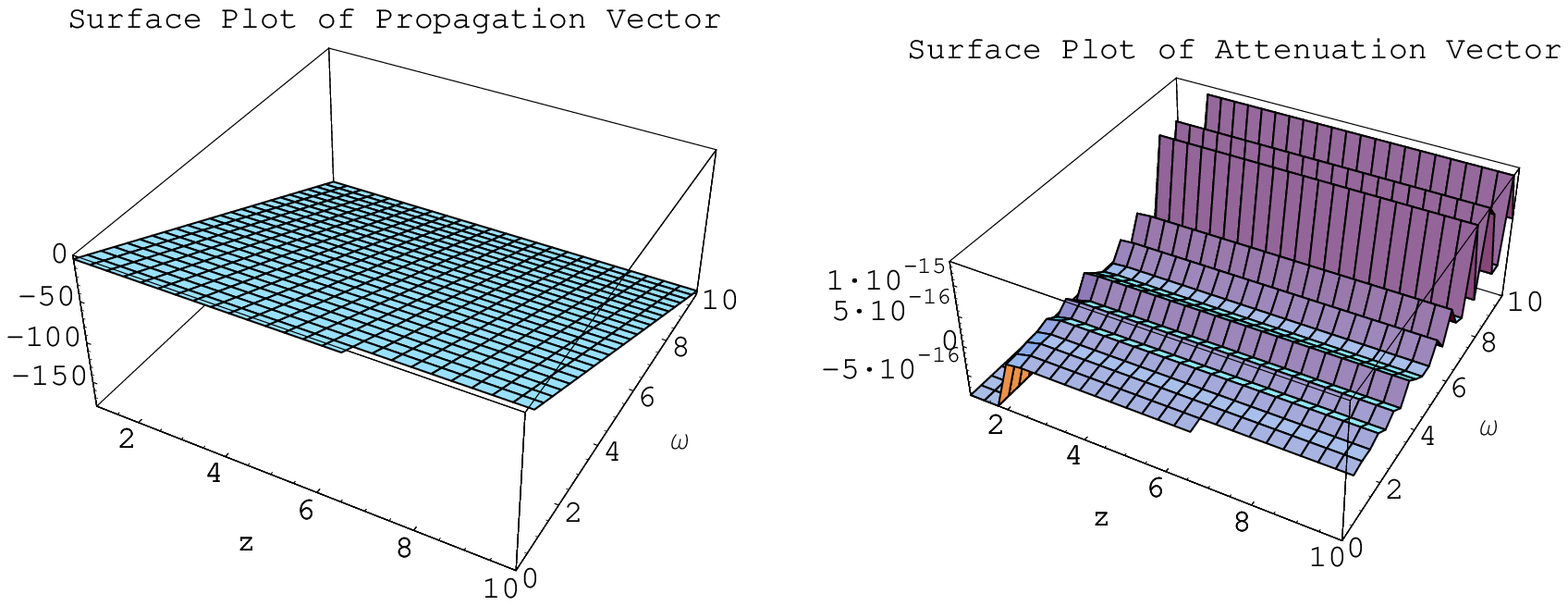,width=0.7\linewidth} \center
\epsfig{file=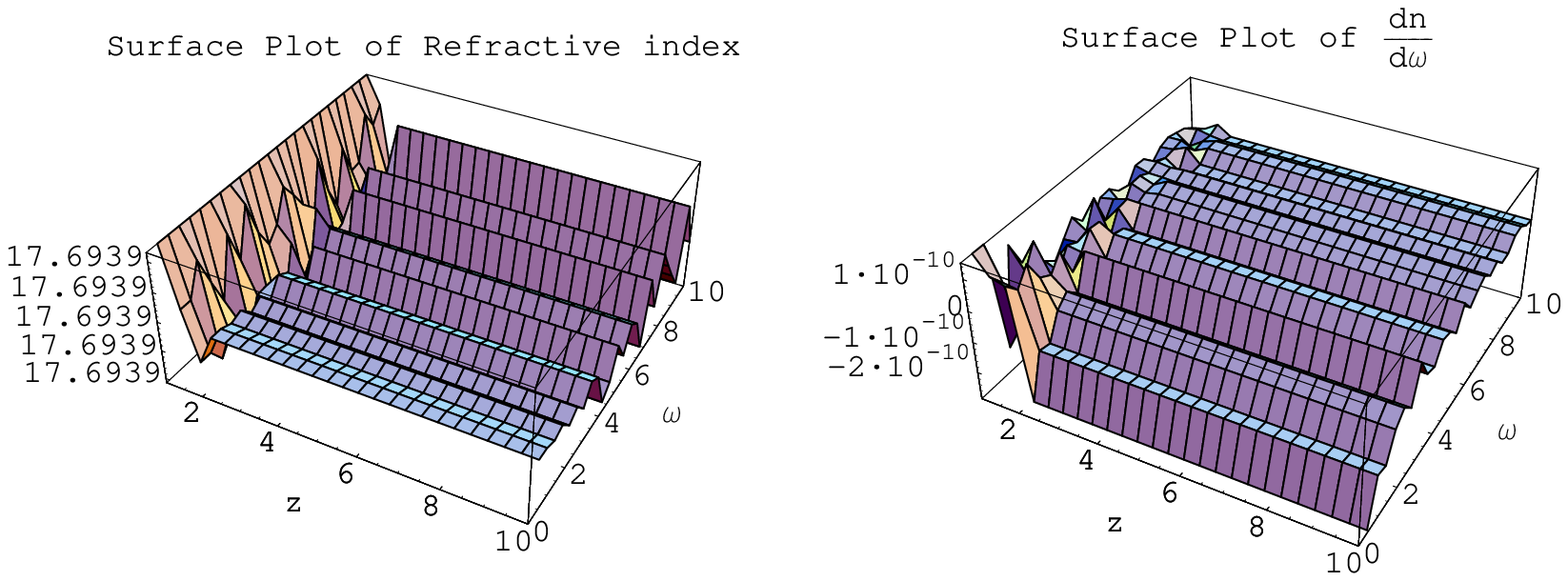,width=0.7\linewidth} \caption{The waves move
towards the event horizon. Most of the region admits the
properties of anomalous dispersion}
\end{figure}
We obtain a complex dispersion relation quartic in $k$
\begin{eqnarray}\label{v7}
&&A_1(z)k^4+A_2(z,\omega)k^3+A_3(z,\omega)k^2
+A_4(z,\omega)k+A_5(z,\omega)\nonumber\\&&+\iota\{A_6(z)k^3+A_7(z,\omega)k^2
+A_8(z,\omega)k+A_9(z,\omega)\}=0.
\end{eqnarray}
The corresponding graphs  are shown in Figures
\textbf{6}-\textbf{9}.

Figure \textbf{6} shows that the propagation vector decreases with
the increase in $\omega$ and $z$. The attenuation vector attains
random values in the region $2.1\leq z\leq10,~0\leq \omega\leq10$
showing that waves damp and grow randomly. In the region $1\leq
z<2.1$, the attenuation vector increases with the increase in $z$.
The variation of refractive index with respect to $\omega$ is
negative in most of the region which shows anomalous dispersion of
waves. There are some points in the region where
$\frac{dn}{d\omega}>0$ and dispersion is normal.
\begin{figure}
\center \epsfig{file=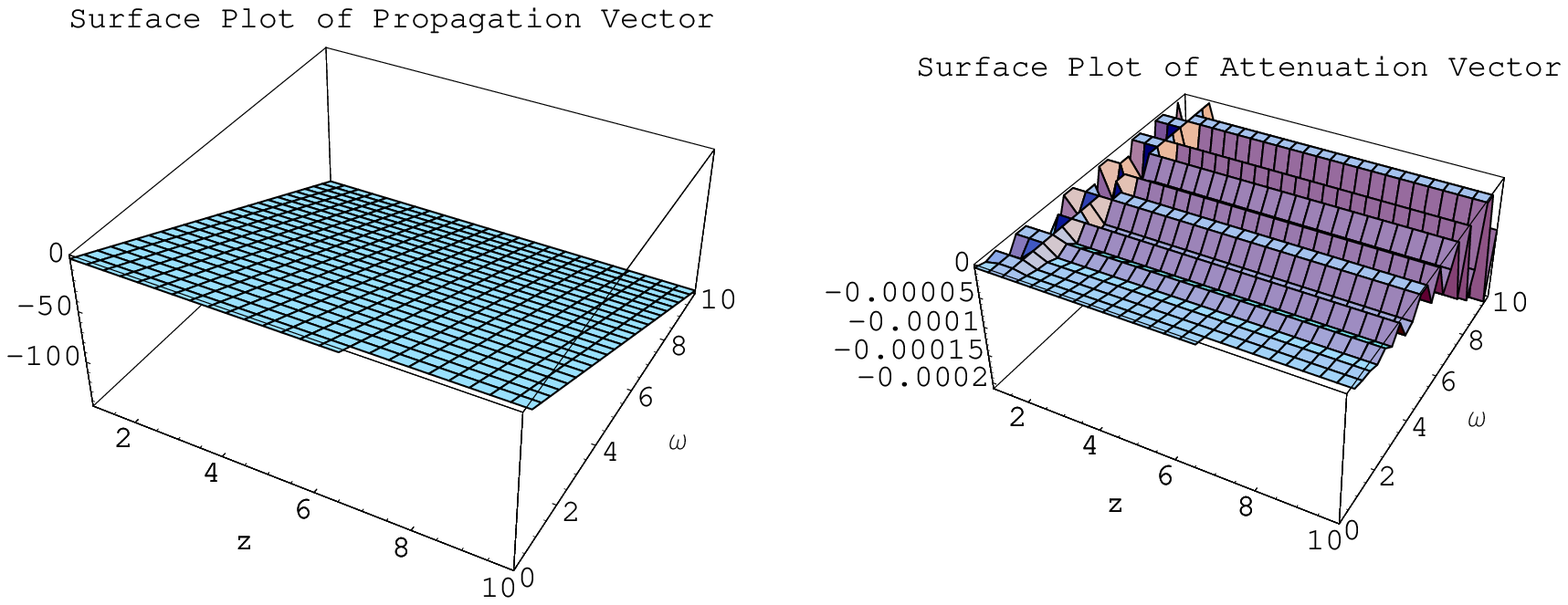,width=0.7\linewidth} \center
\epsfig{file=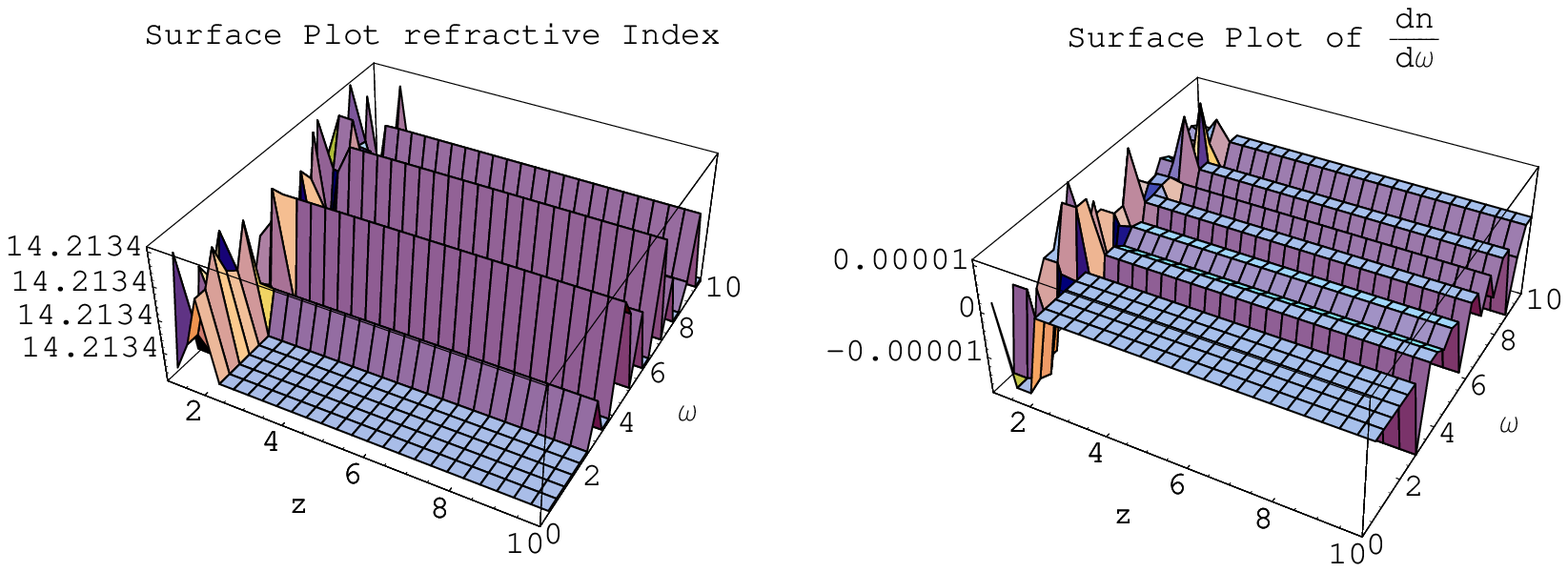,width=0.7\linewidth} \caption{The wave are
directed towards the event horizon. The dispersion is found to be
normal and anomalous randomly}
\end{figure}
Figure \textbf{7} indicates that the attenuation vector takes
negative values. The propagation vector decreases with the
increase in angular frequency and $z$. The attenuation vector
takes random values with the increase in angular frequency. The
refractive index is greater than one and its variation with
respect to angular frequency admit positive values at some points
which indicates normal dispersion of waves.
\begin{figure} \center
\epsfig{file=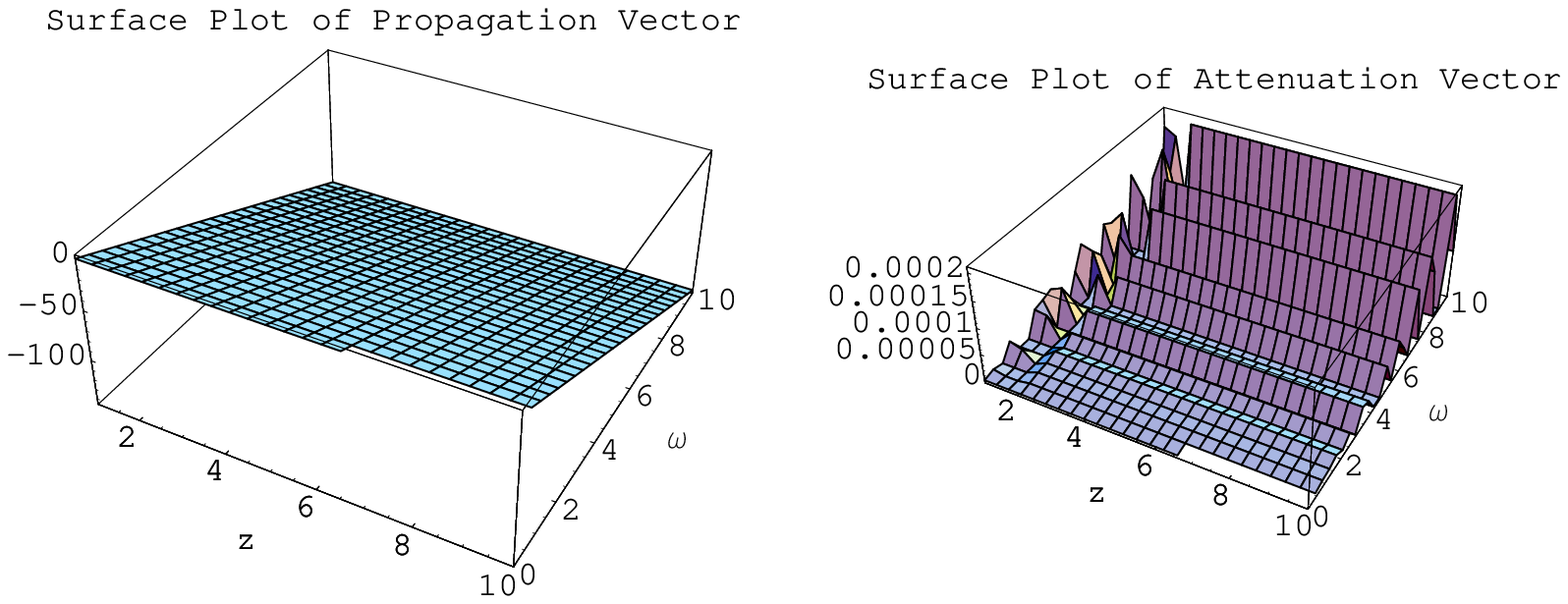,width=0.7\linewidth} \center
\epsfig{file=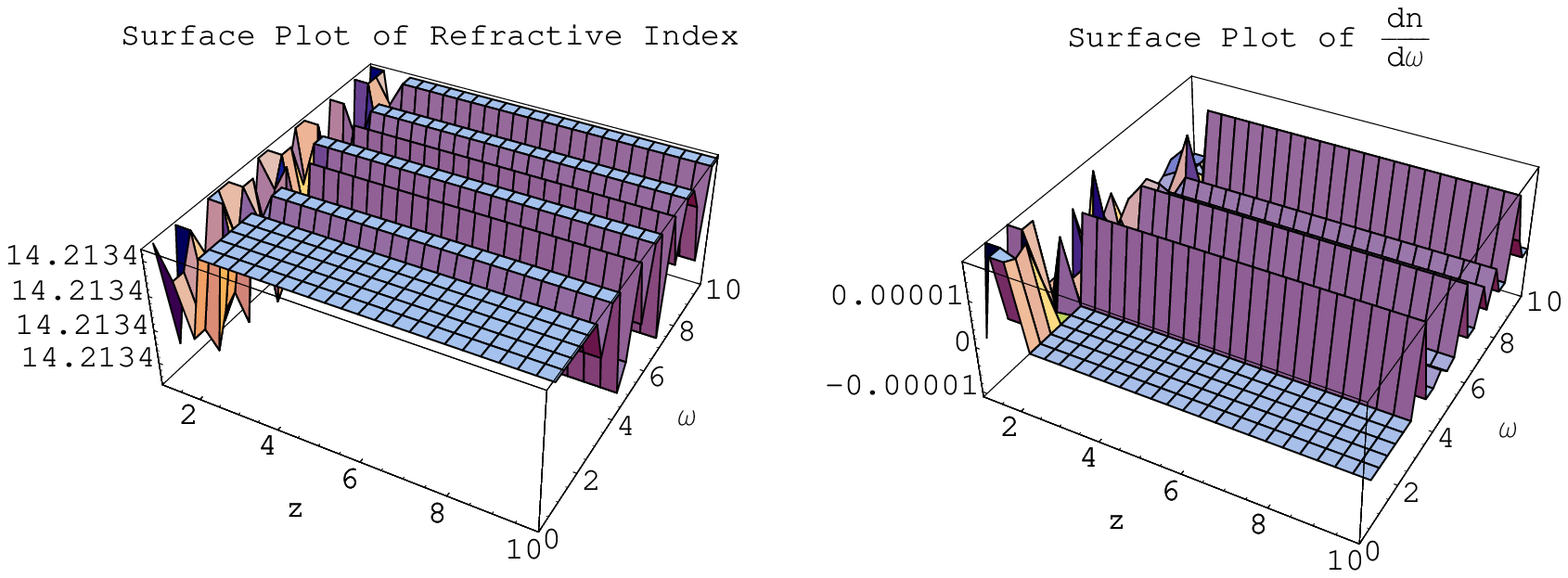,width=0.7\linewidth} \center \caption{The
waves move towards the event horizon. Most of the region admits
normal dispersion}
\end{figure}
Figure \textbf{8} shows that the attenuation vector is positive
throughout the region. The propagation vector decreases with the
increase in angular frequency and $z$. The attenuation vector
randomly increases and decreases with the increase in angular
frequency. It shows that the waves randomly damp and grow with the
increase in angular frequency. The refractive index is greater
than one and $\frac{dn}{d\omega}>0$ in most of the region which
results that the waves disperse normally.
\begin{figure} \center
\epsfig{file=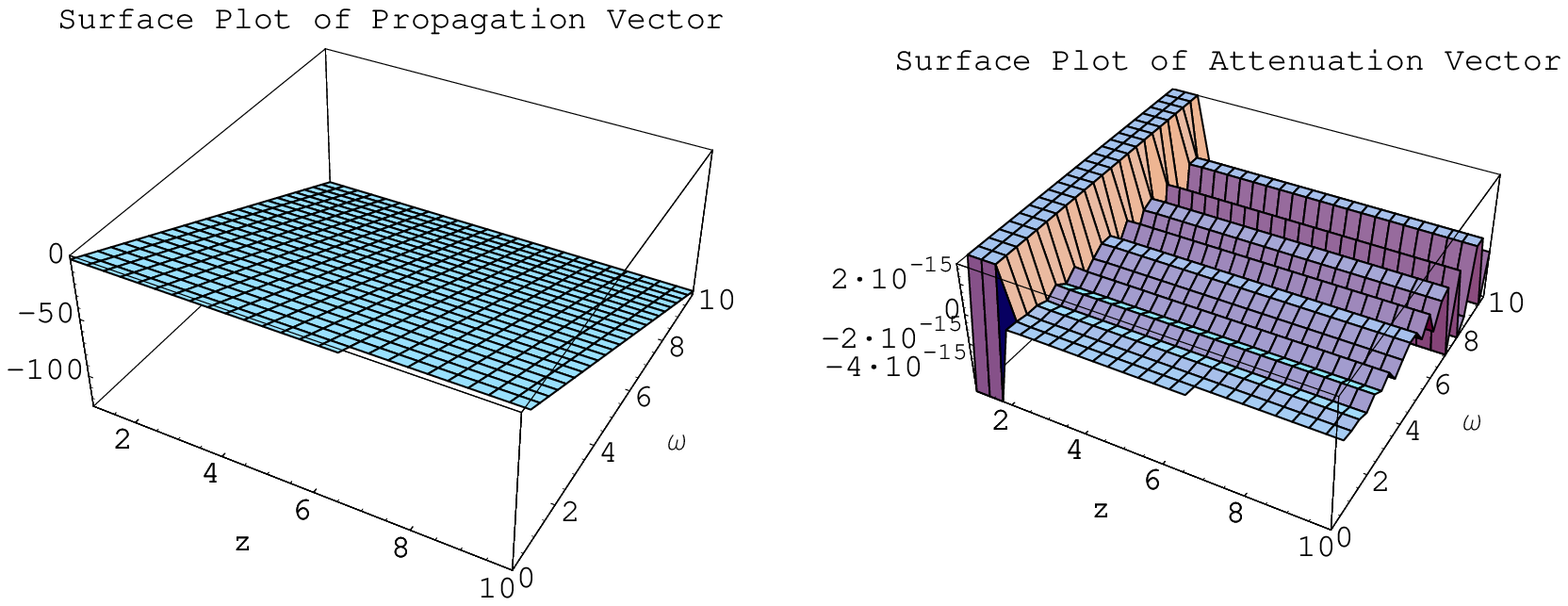,width=0.7\linewidth} \center
\epsfig{file=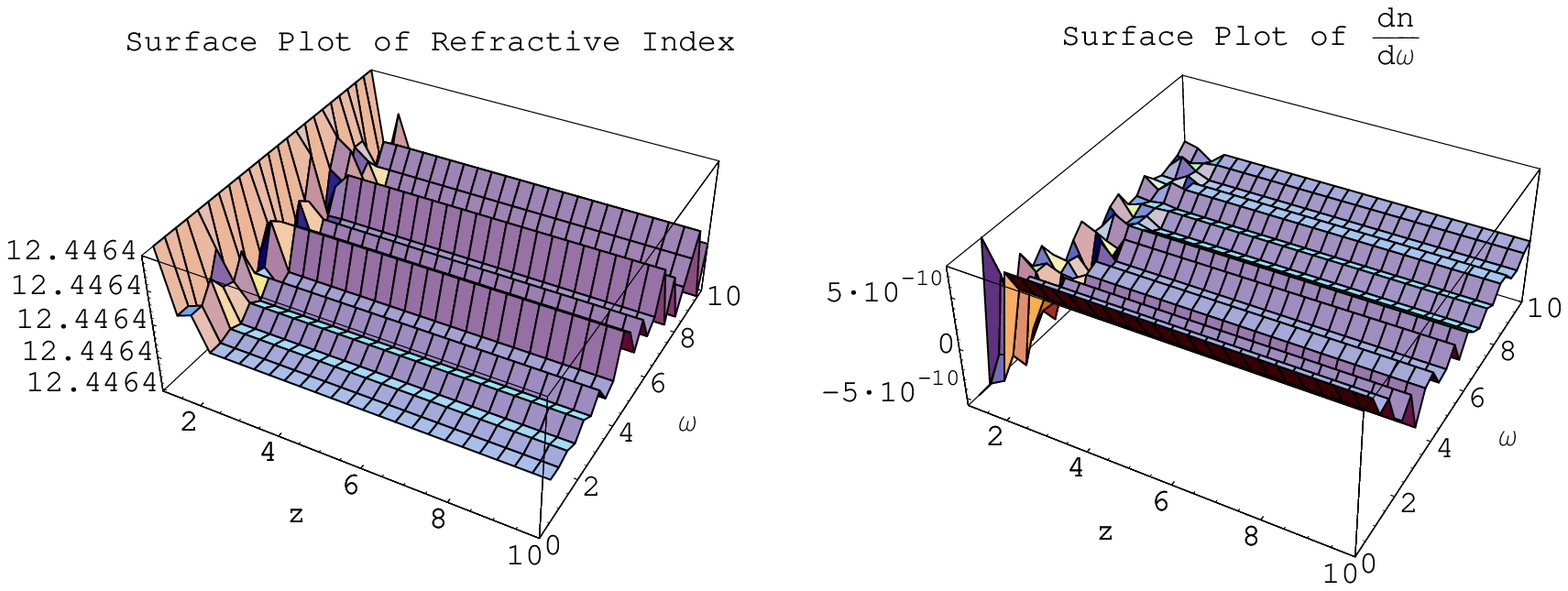,width=0.7\linewidth} \center \caption{The
waves are directed towards the event horizon. Some of the region
admits normal dispersion}
\end{figure}
Figure \textbf{9} shows that the attenuation vector is positive
for the region $1\leq z\leq5,~10^{-5}\leq \omega\leq10$. In a
small region near the event horizon, the attenuation vector
decreases with the increase in $z$ which indicates that the waves
damp as they move towards the event horizon. The refractive index
is greater than one and its variation with respect to angular
frequency is greater than zero at random points which admit normal
dispersion of waves.

In Figures \textbf{6}-\textbf{9}, the propagation vector is
negative in the whole region. Moreover, Figures \textbf{6} and
\textbf{9} admit a small region near the event horizon where the
refractive index increases with the decrease in $z$.

\section{Non-Rotating Background with Isothermal Plasma}

For the isothermal plasma model, the Fourier analyzed perturbed
GRMHD equations (3.3.1)-(3.3.5) of \cite{ut} are given as follows
\begin{eqnarray}\setcounter{equation}{1}\label{h18}
&&-\frac{\iota\omega}{\alpha}c_5=0,\\
\label{h19} &&\iota kc_5=0,\\
\label{h20} &&c_1\{-\rho\iota\omega+\iota k\rho\alpha u-(u\alpha
p)'-\alpha u^2\gamma^2pu'\}+c_2\{-p\iota\omega+\iota kp\alpha
u\nonumber\\
&&+(u\alpha p)'+\alpha
u^2\gamma^2pu'\}+c_3(\rho+p)\{\alpha(1+\gamma^2u^2)\iota
k-\alpha(1-2\gamma^2u^2)\nonumber
\end{eqnarray}
\begin{eqnarray}
&&\times(1+\gamma^2u^2)\frac{u'}{u}-\iota\omega\gamma^2 u\}=0,\\
\label{h21}
&&c_1\rho\gamma^2\{a_z+uu'(1+\gamma^2u^2)\}+c_2\{p\gamma^2\{a_z
+uu'(1+\gamma^2u^2)\}\nonumber\\
&&+\iota
kp+p'\}+c_3(\rho+p)\gamma^2[(1+\gamma^2u^2)(-\frac{\iota\omega}{\alpha}
+\iota ku)+\{u'(1+\gamma^2u^2)\nonumber\\
&&\times(1+4\gamma^2u^2)+2\gamma^2ua_z\}]=0,\\
\label{h22}&&c_1\rho\gamma^2\{-\frac{\iota\omega}{\alpha}+u(a_z
+\gamma^2uu')\}
+c_2p\{-\frac{\iota\omega}{\alpha}(\gamma^2-1)+\gamma^2u(a_z\nonumber\\
&&+\gamma^2uu')\}+c_3(\rho+p)\gamma^2\{-\frac{2\iota\omega}{\alpha}
\gamma^2u+\gamma^2u^2\iota k+(2\gamma^2uu'+a_z)\nonumber\\
&&\times(1+2\gamma^2 u^2)\}=0.
\end{eqnarray}
Equations (\ref{h18}) and (\ref{h19}) show that $c_5=0$ which
means that there are no perturbations in magnetic field.

\subsection{Numerical Solutions}

In order to find the numerical solutions we assume the same time
lapse and $z$-component of fluid velocity as in Section
\textbf{4.1}. When we substitute these values with the assumption
of stiff fluid, i.e., $\rho=p$, the mass conservation law in three
dimensions gives $\rho=-\frac{1}{2u}=p$. The complex dispersion
relation is of the form
\begin{eqnarray}\label{v9}
&&A_1(z)k^2+A_2(z,\omega)k+A_3(z,\omega)+\iota\{A_4(z)k^3+A_5(z,\omega)k^2
\nonumber\\
&&+A_6(z,\omega)k+A_7(z,\omega)\}=0.
\end{eqnarray}
This gives three complex values of $k$ shown in Figures
\textbf{10}-\textbf{12}.

\begin{figure} \center
\epsfig{file=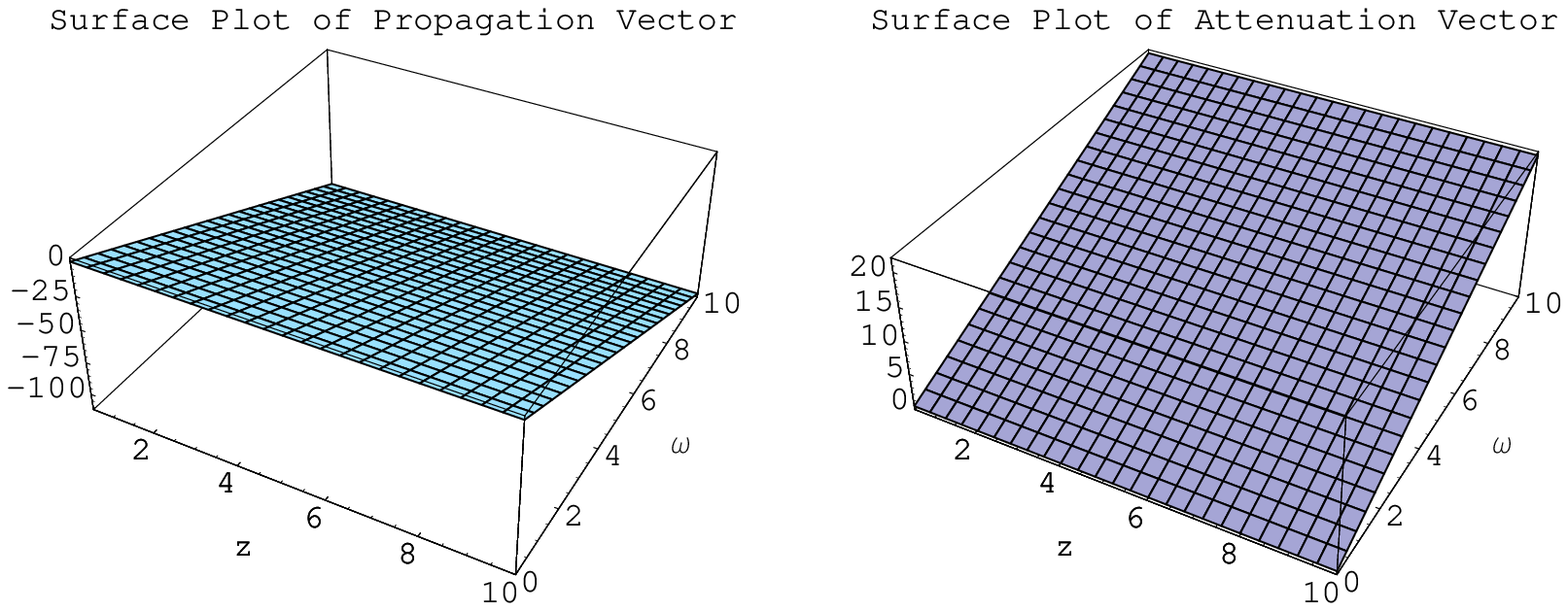,width=0.7\linewidth}
\center\epsfig{file=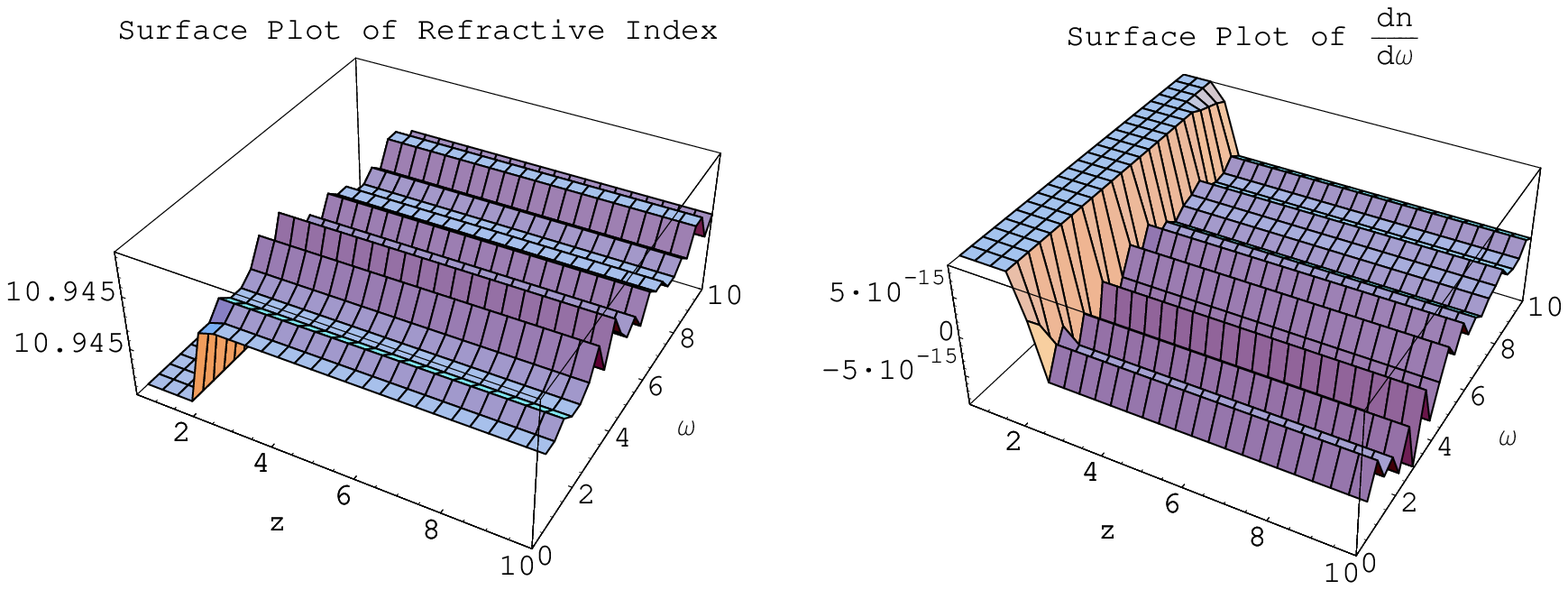,width=0.7\linewidth} \caption{The
waves move towards the event horizon. A small region near the
event horizon admits normal dispersion}
\end{figure}
Figure \textbf{10} indicates that the propagation decreases while
the attenuation vector increases with the increase in angular
frequency. The refractive index is greater than one and its
variation with respect to $\omega$ is greater than zero in the
region $0.5\leq z\leq2$. Rest of the region admits points of
normal as well as anomalous dispersion.
\begin{figure}
\center \epsfig{file=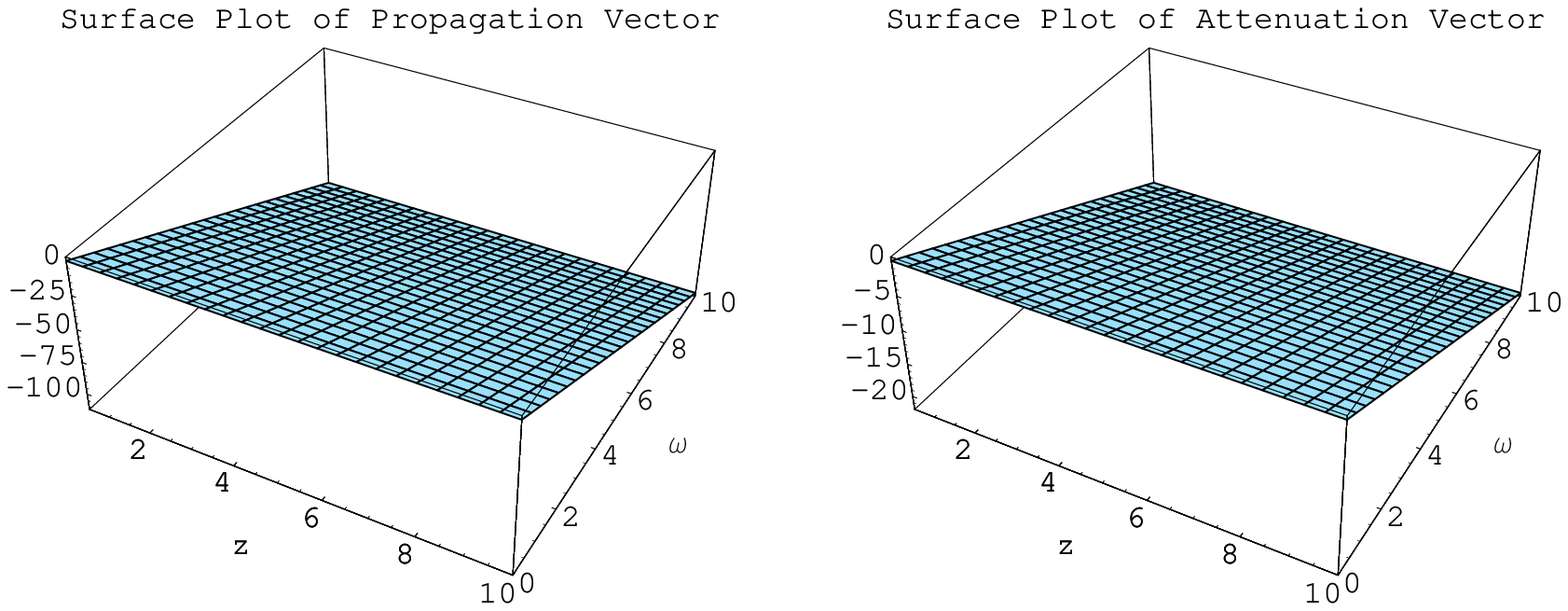,width=0.7\linewidth}
\center\epsfig{file=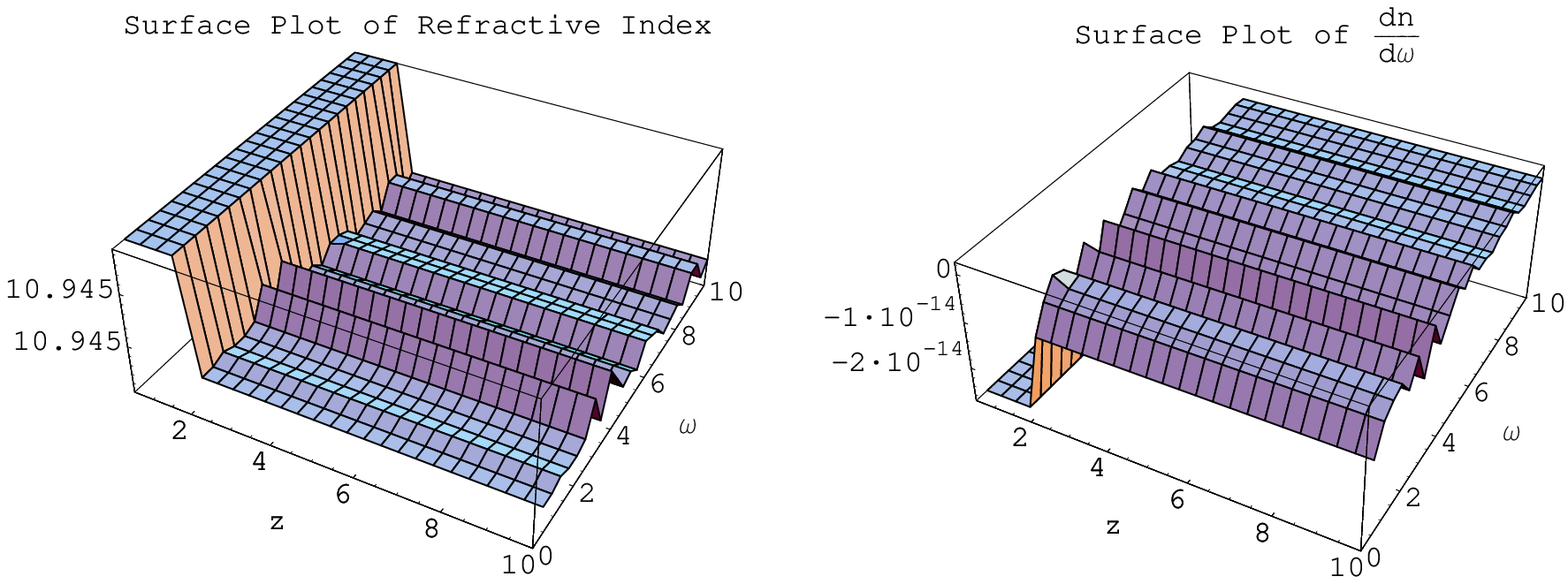,width=0.7\linewidth} \caption{The
waves are directed towards the event horizon. Dispersion is found
to be anomalous}
\end{figure}
Figure \textbf{11} shows that the propagation vector decreases
with the increase in angular frequency. The attenuation vector is
also negative and it decreases with the increase in angular
frequency and $z$. It shows that the waves grow with the increase
in angular frequency and $z$. The refractive index is greater than
one and its variation with respect to $\omega$ is negative which
indicates that the region admits anomalous dispersion.
\begin{figure} \center
\epsfig{file=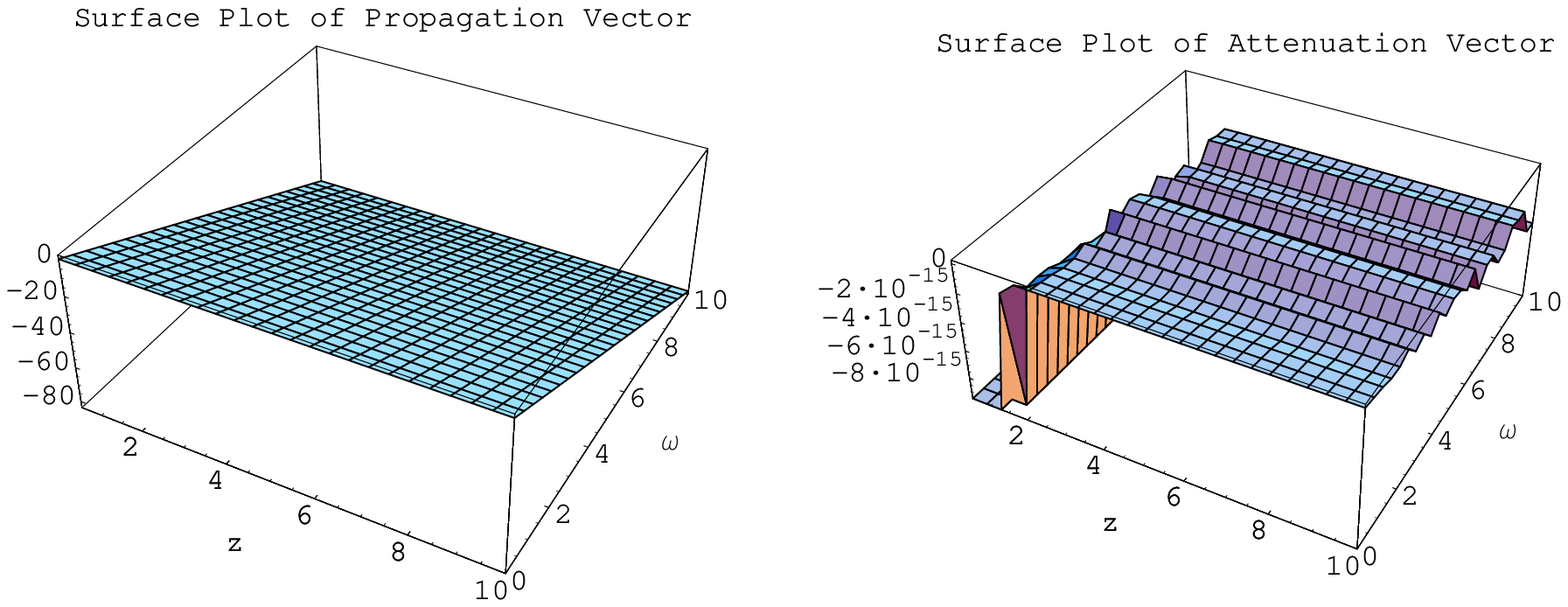,width=0.7\linewidth}
\center\epsfig{file=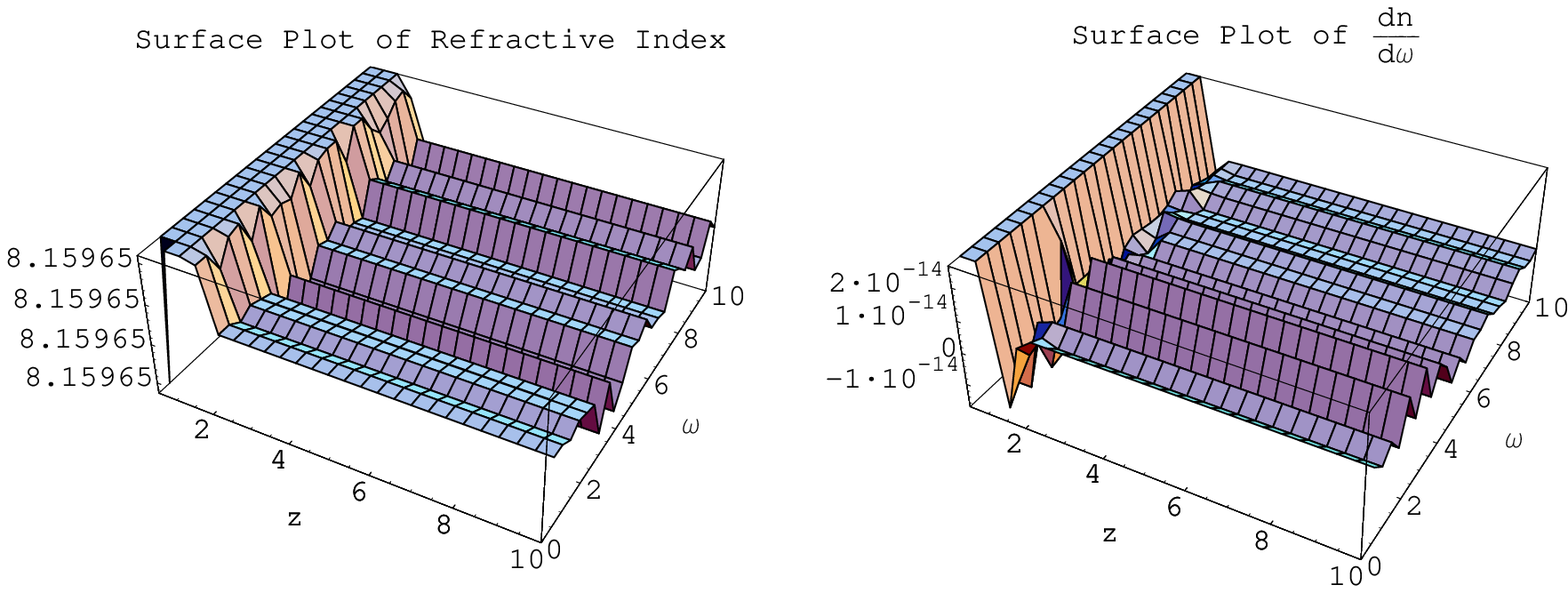,width=0.7\linewidth} \caption{The
waves move toward the event horizon. The dispersion is normal in a
small region near the event horizon}
\end{figure}
Figure \textbf{12} shows that the attenuation vector admits
negative values. The propagation vector decreases while the
attenuation vector shows random increase and decrease with the
increase in angular frequency. The attenuation vector decreases
near the event horizon which indicates that the waves grow near
the event horizon. The refractive index is greater than one. The
change in refractive index with respect to angular frequency is
greater than zero in the region $0.5\leq z\leq3, ~
0\leq\omega\leq10$ which indicates normal dispersion.

The propagation vector takes negative values for Figures
\textbf{10}-\textbf{12}. In a small region near the event horizon,
the refractive index increases with the increase in $z$ for
Figures \textbf{11}-\textbf{12}.

\section{Rotating Non-magnetized Background with Isothermal Plasma}

The Fourier analyzed GRHD equations for isothermal plasma are
given by Eqs.(3.4.1)-(3.4.4) of \cite{ut}.
\begin{eqnarray}\setcounter{equation}{1}\label{j6}
&&c_1\{(-\iota\omega+\iota\alpha ku)\rho-\alpha'up-\alpha
u'p-\alpha up'-\alpha\gamma^2up(VV'+uu')\}\nonumber
\end{eqnarray}
\begin{eqnarray}
&&+c_2\{(-\iota\omega+\iota\alpha ku)p+\alpha'up+\alpha u'p+\alpha
up'+\alpha\gamma^2up(VV'+uu')\}\nonumber\\
&&+c_3(\rho+p)[-\iota\omega\gamma^2u+\iota
k\alpha(1+\gamma^2u^2)-\alpha\{(1-2\gamma^2u^2)(1+\gamma^2u^2)\frac{u'}{u}
\nonumber\\
&&-2\gamma^4u^2VV'\}]+c_4(\rho+p)[\gamma^2V(-\iota\omega+\iota
k\alpha u)+\alpha\gamma^2u\{(1+2\gamma^2V^2)V'\nonumber\\
&&+2\gamma^2uVu'\}]=0,\\
\label{j7}&&c_1\rho\gamma^2u\{(1+\gamma^2V^2)V'+\gamma^2uVu'\}
+c_2p\gamma^2u\{(1+\gamma^2V^2)V'+\gamma^2uVu'\} \nonumber\\
&&+c_3(\rho+p)\left[\gamma^2uV\left(\frac{-\iota\omega}{\alpha}+\iota
k\alpha\right) +\{(1+2\gamma^2V^2)(1+2\gamma^2u^2)-\gamma^2V^2\}V'
\right.\nonumber\\
&&\left.+2\gamma^2uVu'(1+2\gamma^2u^2)\right]
+c_4(\rho+p)\gamma^2\left[(1+\gamma^2V^2)
\left(\frac{-\iota\omega}{\alpha}+\iota
ku\right)\right.\nonumber\\&&\left.
+\gamma^2u\{(1+4\gamma^2u^2)uu'+4(1+\gamma^2V^2)VV'\}\right]=0,\\
\label{j8}
&&c_1\rho\gamma^2\{a_z+uu'(1+\gamma^2u^2)+\gamma^2u^2VV'\}+
c_2[p\gamma^2\{a_z+uu'(1+\gamma^2u^2)\nonumber\\
&&+\gamma^2u^2VV'\}+p'+\iota
kp]+c_3(\rho+p)\gamma^2\left[(1+\gamma^2u^2)
\left(\frac{-\iota\omega}{\alpha}+\iota
ku\right)\right.\nonumber\\
&&\left.+u'(1+\gamma^2u^2)(1+4\gamma^2u^2)+2u\gamma^2\{(1+2\gamma^2u^2)VV'
+a_z\}\right]\nonumber\\
&&+c_4(\rho+p)\gamma^4\left[uV\left(\frac{-\iota\omega}{\alpha}+\iota
ku\right)
+u^2V'(1+4\gamma^2u^2)\right.\nonumber\\
&&\left.+2V\{(1+2\gamma^2u^2)uu'+a_z\}\right]=0,\\
\label{j9}&&c_1\rho\gamma^2\left[\frac{-\iota\omega}{\alpha}
+u\{a_z+\gamma^2(VV'+uu')\}\right]+c_2p\left[\frac{-\iota\omega}{\alpha}(\gamma^2-1)
+u\gamma^2\{a_z\right.\nonumber\\
&&\left.+\gamma^2(VV'+u u') \}\right]+c_3(\rho+p)\gamma^2
\left\{\gamma^2u\left(-\frac{2\iota\omega}{\alpha}+\iota
ku\right)+(a_z+\gamma^2uu')\right.\nonumber\\
&&\left.(1+2\gamma^2u^2)+\gamma^2VV'(1+4\gamma^2u^2)\right\}
+c_4(\rho+p)\gamma^2\left\{\gamma^2V\left(-\frac{2\iota\omega}{\alpha}
+\iota
ku\right)\right.\nonumber\\
&&\left.+\gamma^2uV'(1+4\gamma^2V^2)+2\gamma^2uV(a_z+2\gamma^2uu')\right\}=0.
\end{eqnarray}

\subsection{Numerical Solutions}

Using the same assumptions as given in Section \textbf{5.1}, we
obtain a complex dispersion relation of the form
\begin{eqnarray}\label{v10}
&&A_1(z)k^4+A_2(z,\omega)k^3+A_3(z,\omega)k^2+A_4(z,\omega)k+A_5(z,\omega)\nonumber\\
&&+\iota\{A_6(z)k^3+A_7(z,\omega)k^2+A_8(z,\omega)k+A_9(z,\omega)\}=0.
\end{eqnarray}
The four complex values of $k$ are represented in Figures
\textbf{13}-\textbf{16}.
\begin{figure}
\center \epsfig{file=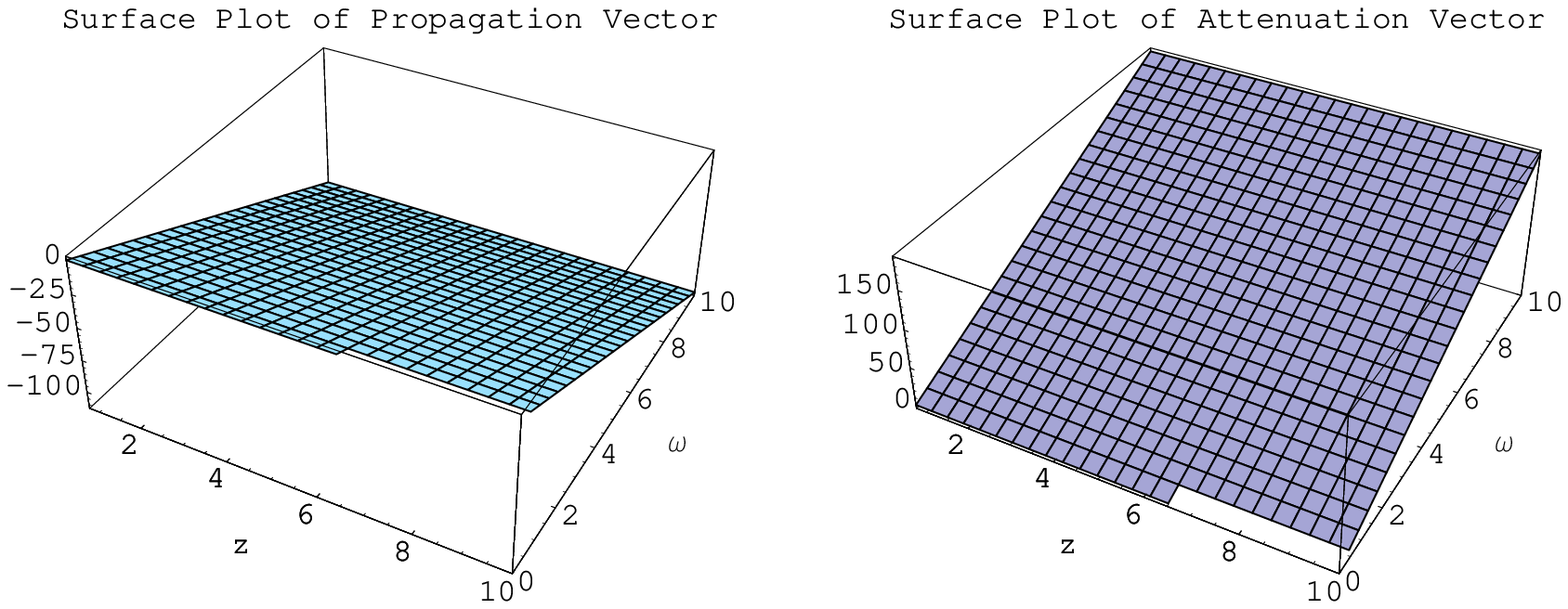,width=0.7\linewidth}
\center\epsfig{file=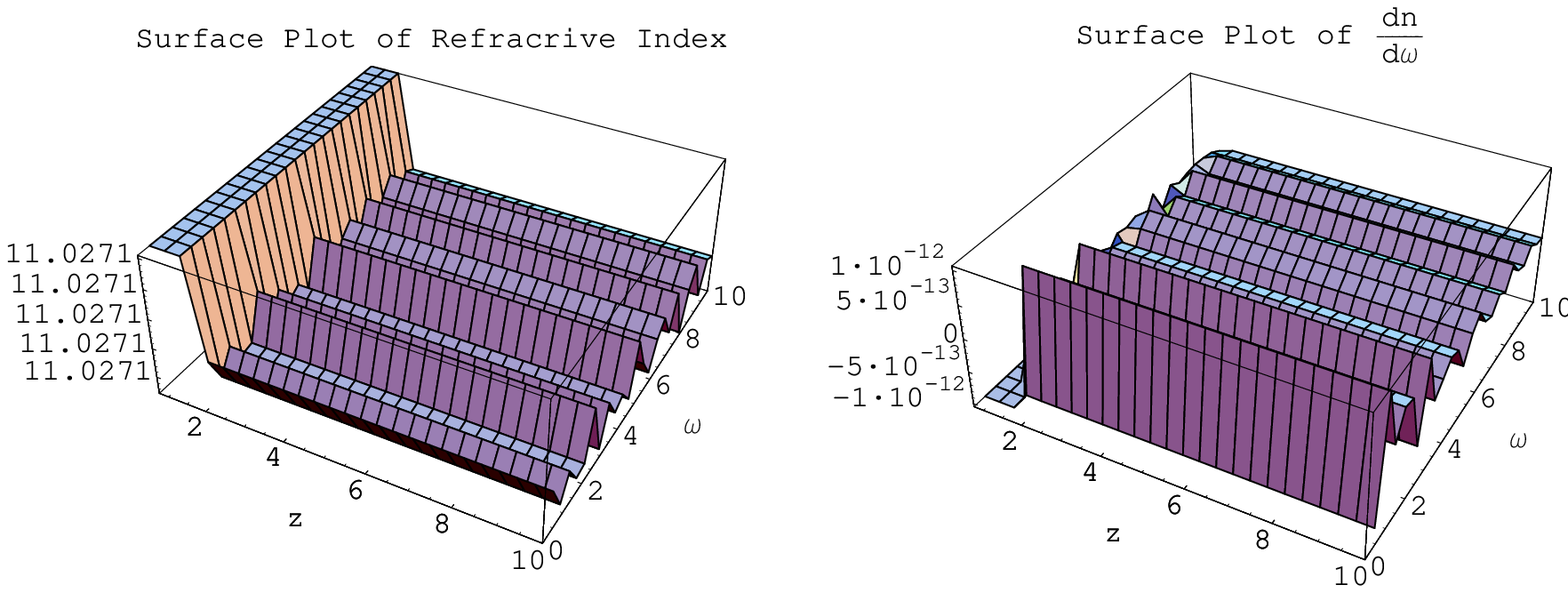,width=0.7\linewidth} \caption{The
waves move towards the event horizon. A small region near the
event horizon admits anomalous dispersion}
\end{figure}

Figure \textbf{13} shows that the propagation vector takes
negative values. The propagation vector decreases with the
increase in angular frequency and $z$. The attenuation vector
increases with the increase in angular frequency. The refractive
index is greater than one and its variation with respect to
$\omega$ is less than zero in the region $0.75\leq z\leq4.5, ~
0\leq\omega\leq10$ which leads to  anomalous dispersion. Random
points of normal and anomalous dispersion are found otherwise.
\begin{figure}
\center \epsfig{file=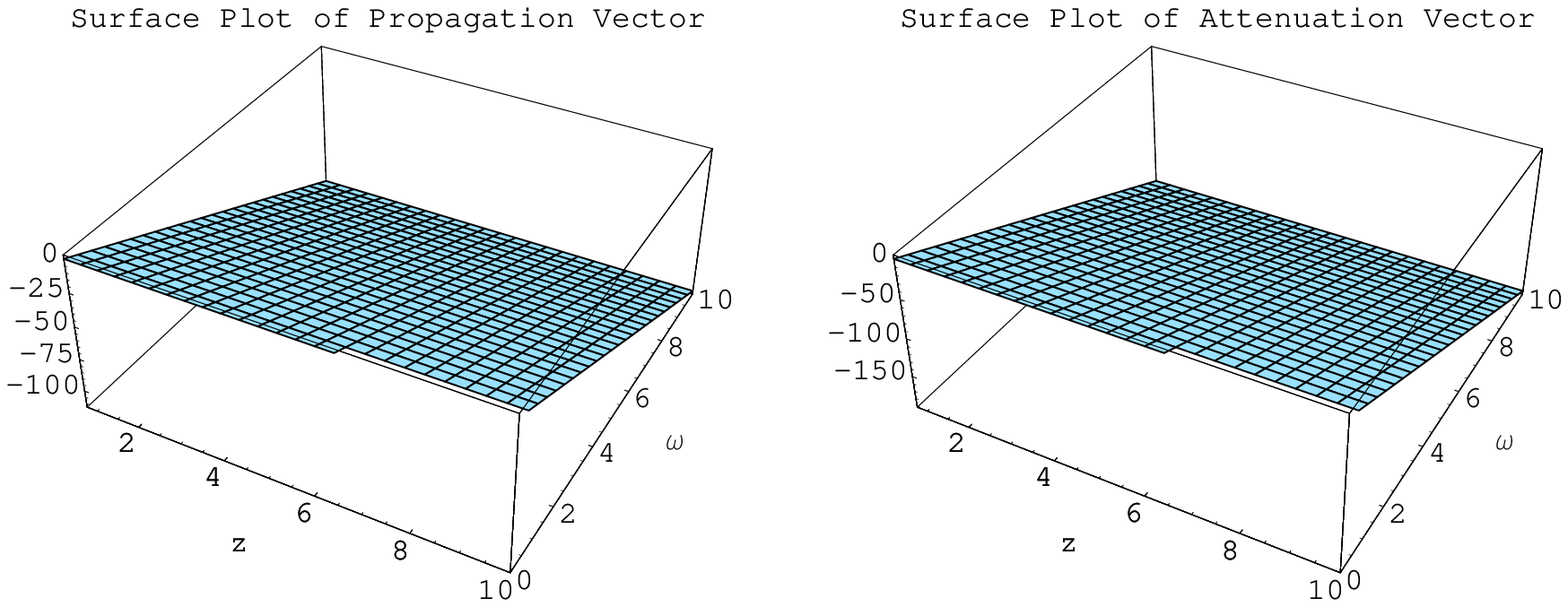,width=0.7\linewidth}
\center\epsfig{file=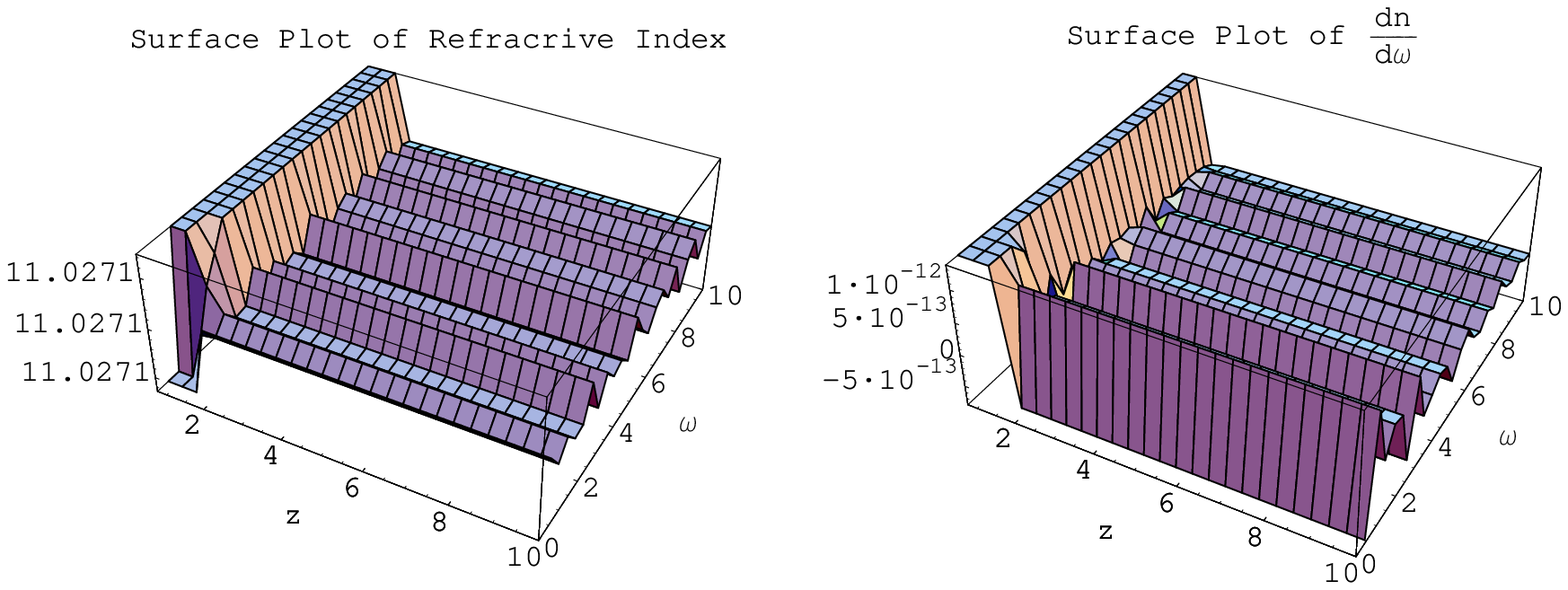,width=0.7\linewidth} \caption{The
waves are directed towards the event horizon. Normal dispersion of
waves is found near the event horizon}
\end{figure}
Figure \textbf{14} indicates that the propagation and attenuation
vectors decrease with the increase in $z$ and $\omega$. It shows
that the waves grow with the increase in angular frequency and
$z$. The refractive index is greater than one and its variation
with respect to $\omega$ is positive in the region $0.75\leq
z\leq4, ~ 0\leq\omega\leq10$ which indicates normal dispersion of
waves. Random points of normal and anomalous dispersion lies in
rest of the region.
\begin{figure}
\center \epsfig{file=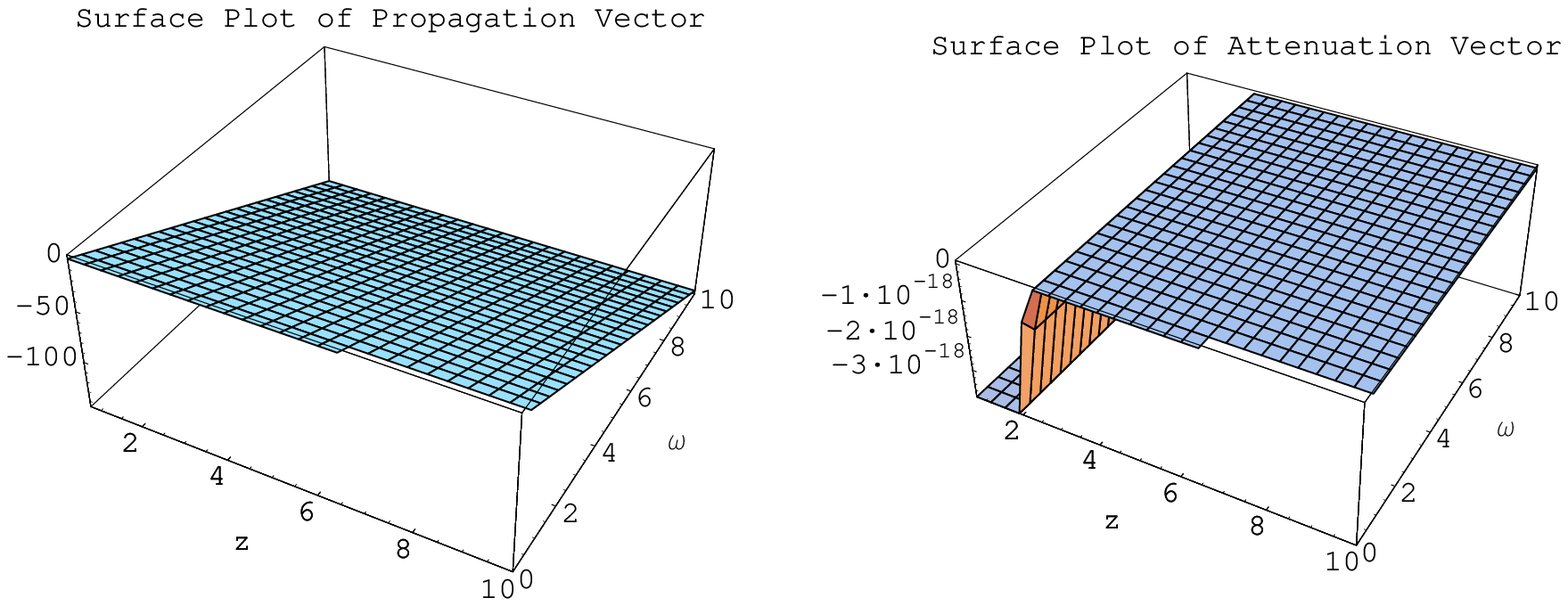,width=0.7\linewidth}
\center\epsfig{file=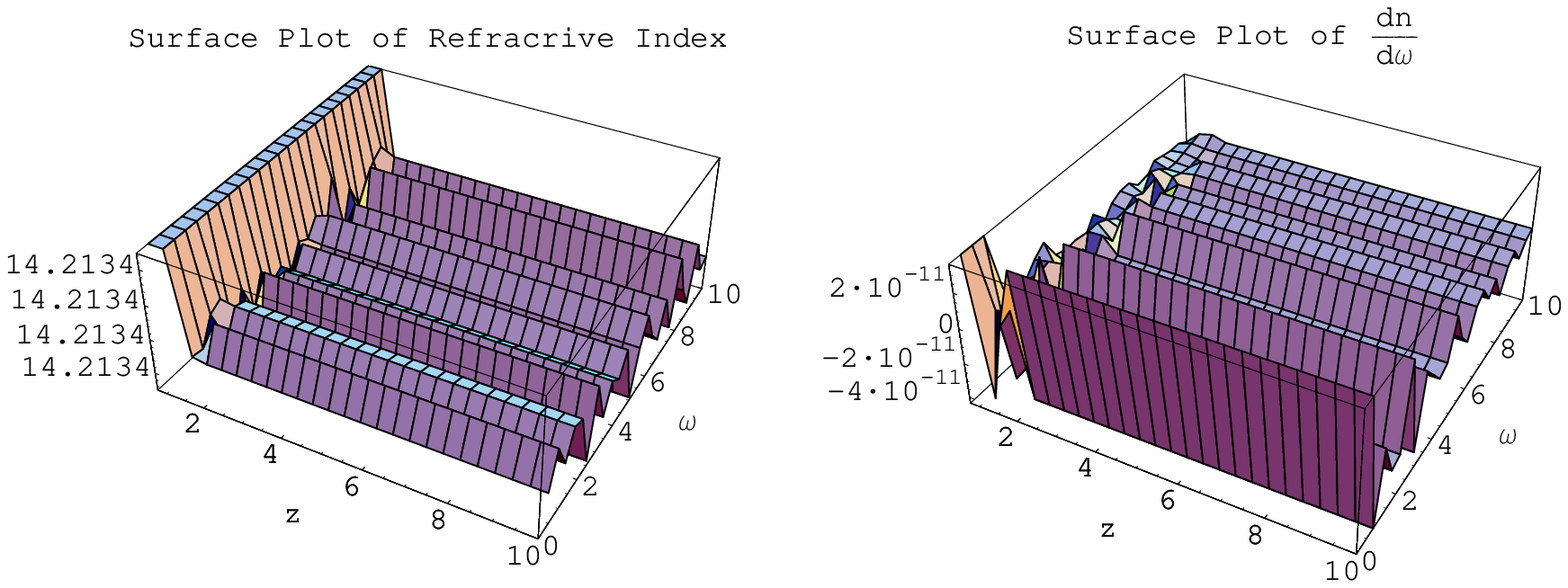,width=0.7\linewidth} \caption{The
waves are directed towards the event horizon. Dispersion is found
to be normal at random points}
\end{figure}
Figure \textbf{15} shows that the propagation vector decreases
with the increase in angular frequency and $z$. The attenuation
vector decreases as $z$ decreases which indicates that the waves
grow as they move towards the event horizon. The refractive index
is greater than one. The quantity $\frac{dn}{d\omega}>0$ at random
points which leads to normal dispersion of waves at those points.
\begin{figure}
\center \epsfig{file=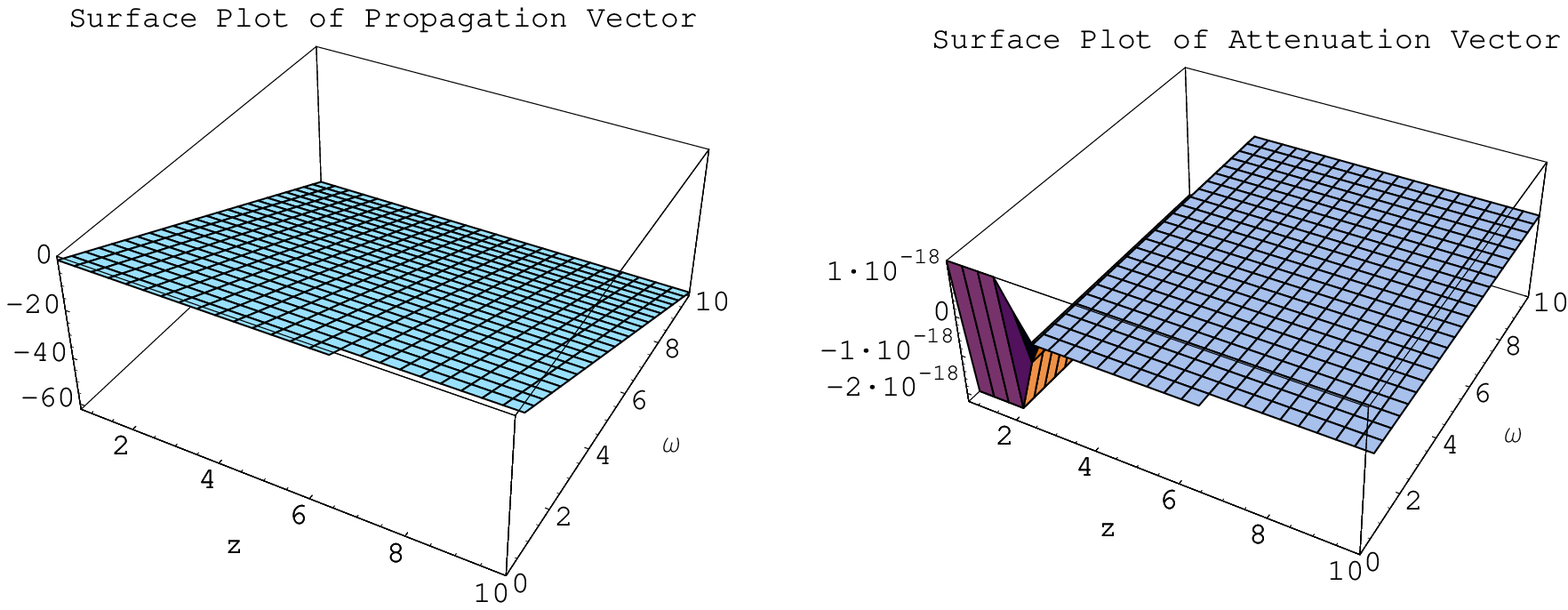,width=0.7\linewidth} \center
\epsfig{file=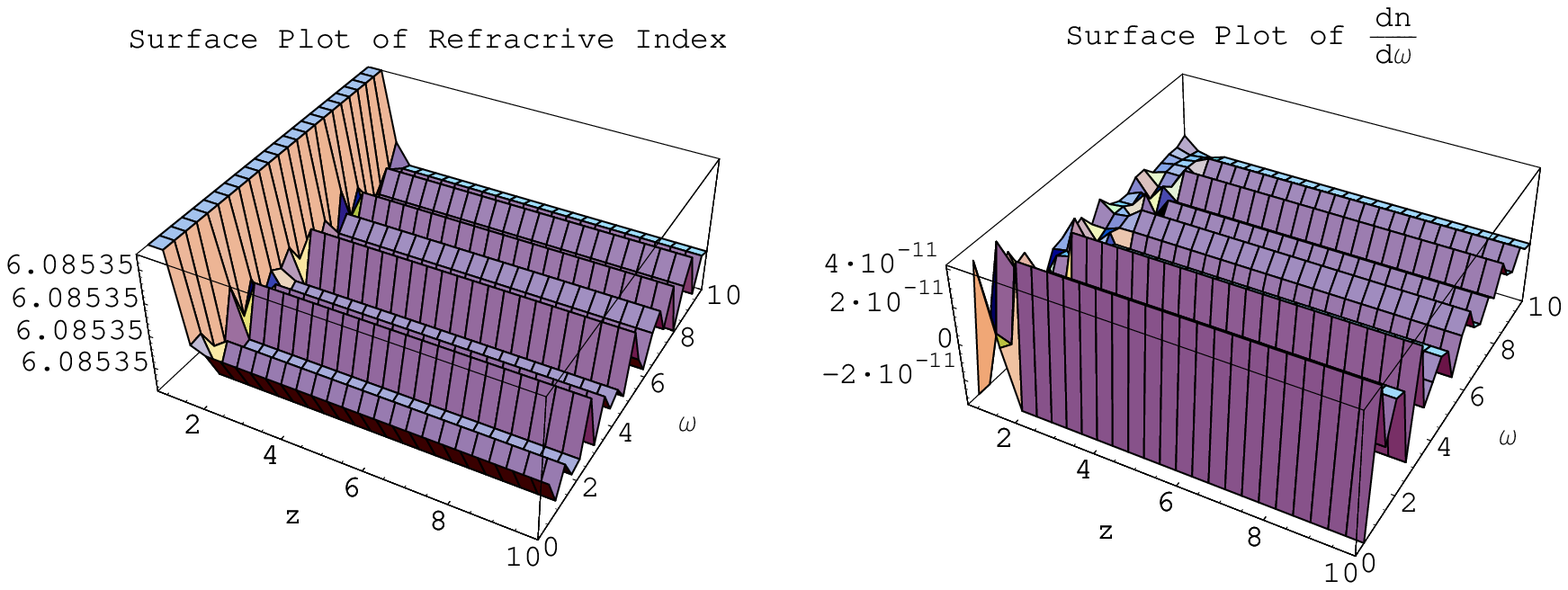,width=0.7\linewidth} \caption{The waves move
towards the event horizon. Dispersion is found to be normal at
random points}
\end{figure}
In Figure \textbf{16}, the attenuation vector decreases as $z$
decreases in the region except for $\omega=0$ which shows the
growth of waves as they move towards the event horizon. The
refractive index is greater than one and its variation with
respect to $\omega$ is greater than zero at random points which
shows that the waves disperse normally at random points.

In Figures \textbf{13}-\textbf{16}, the propagation vector admits
negative values which indicates that the waves move towards the
event horizon. The refractive index increases in a small region
near the event horizon for all these figures.

\section{Rotating Magnetized Background with\\Isothermal Plasma}

This is the most general and physical case considered to
investigate the wave properties, i.e., we take rotating isothermal
plasma which is obviously highly magnetized. The respective
Fourier analyzed perturbed GRMHD equations are given by
Eqs.(3.5.1)-(3.5.7) of \cite{ut}
\begin{eqnarray}\setcounter{equation}{1}
\label{j19} &&c_4(\alpha'+\iota
k\alpha)-c_3\{(\alpha\lambda)'+\iota k\alpha\lambda\}+c_5(\alpha
V)'-c_6\{(\alpha u)'-\iota\omega\nonumber\end{eqnarray}
\begin{eqnarray}
&&+\iota ku\alpha\}=0,\\\label{j20}
&&c_5(\frac{-\iota\omega}{\alpha}+\iota k\alpha)=0,\\
\label{j21}
&&c_5\iota k=0,\\
\label{j22} &&c_1\{(-\iota\omega+\iota\alpha ku)\rho-\alpha'up
-\alpha u'p-\alpha up'-\alpha\gamma^2up(VV'+uu')\}\nonumber\\
&&+c_2\{(-\iota\omega+\iota\alpha ku)p+\alpha'up+\alpha u'p+\alpha
up'+\alpha\gamma^2up(VV'+uu')\}\nonumber\\
&&+c_3(\rho+p)[-\iota\omega\gamma^2u+\iota k\alpha(1+\gamma^2u^2)
-\alpha\{(1-2\gamma^2u^2)(1+\gamma^2u^2)\frac{u'}{u}\nonumber
\\&&-2\gamma^4u^2VV'\}]+c_4(\rho+p)[\gamma^2V(-\iota\omega+\iota
k\alpha u)\nonumber\\
&&+\alpha\gamma^2u\{(1+2\gamma^2V^2)V'+2\gamma^2uVu'\}]=0,\\\label{j23}
&&c_1\gamma^2\rho u\{(1+\gamma^2V^2)V'+\gamma^2uVu'\}
+c_2\gamma^2p
u\{(1+\gamma^2V^2)V'+\gamma^2uVu'\}\nonumber\\
&&+c_3[-\{(\rho+p)\gamma^4uV-\frac{\lambda
B^2}{4\pi}\}\frac{\iota\omega}{\alpha}+\{(\rho+p)\gamma^4uV+\frac{\lambda
B^2}{4\pi}\}\iota
ku\nonumber\\
&&+(\rho+p)\gamma^2\{(1+2\gamma^2u^2)(1+2\gamma^2V^2)
-\gamma^2V^2\}V'+2(\rho+p)\gamma^4(1\nonumber\\
&&+2\gamma^2u^2)uVu'+\frac{B^2u}{4\pi\alpha}(\alpha\lambda)']
+c_4[-\{(\rho+p)\gamma^2(1+\gamma^2V^2)+\frac{B^2}{4\pi}\}
\frac{\iota\omega}{\alpha}\nonumber\\
&&+\{(\rho+p)\gamma^2(1+\gamma^2V^2)-\frac{B^2}{4\pi}\}\iota
ku+(\rho+p)\gamma^4u \{(1+4\gamma^2V^2)uu'\nonumber\\
&&+4VV'(1+\gamma^2V^2)\}-\frac{B^2u\alpha'}{4\pi\alpha}]
-\frac{B^2}{4\pi}c_6\{(1-u^2)\iota
k+\frac{\alpha'}{\alpha}(1-u^2)\nonumber\\
&&- uu'\}=0,\\
\label{j24} &&c_1\rho\gamma^2\{a_z+(1+\gamma^2u^2)u\acute{u}
+\gamma^2u^2VV'\}+c_2[p\gamma^2\{a_z+(1+\gamma^2u^2)uu'\nonumber\\
&&+\gamma^2u^2VV'\}+\iota
kp+p']+c_3[-\{(\rho+p)\gamma^2(1+\gamma^2u^2)
+\frac{\lambda^2B^2}{4\pi}\}\frac{\iota\omega}{\alpha}\nonumber\\
&&+\{(\rho+p)\gamma^2(1+\gamma^2u^2)
-\frac{\lambda^2B^2}{4\pi}\}\iota
ku+\{(\rho+p)\gamma^2\{u'(1+\gamma^2u^2)\nonumber\\
&&(1+4\gamma^2u^2)
+2u\gamma^2(a_z+(1+2\gamma^2u^2)VV')\}-\frac{\lambda
B^2u}{4\pi\alpha}(\alpha\lambda)'\}]\nonumber\\
&&+c_4[-((\rho+p)\gamma^4uV-\frac{\lambda
B^2}{4\pi})\frac{\iota\omega}{\alpha}+\{(\rho+p)\gamma^4uV+\frac{\lambda
B^2}{4\pi}\}\iota ku\nonumber
\\&&+\{(\rho+p)\gamma^4\{u^2V'(1+4\gamma^2V^2)+2V(a_z
+(1+2\gamma^2u^2)uu')\}+\frac{\lambda
B^2\alpha'u}{4\pi\alpha}\}]\nonumber\\&&+\frac{B^2}{4\pi}c_6[\lambda(1-u^2)\iota
k+\lambda\frac{\alpha'}{\alpha}(1-u^2)-\lambda
uu'+\frac{(\alpha\lambda)'}{\alpha}]=0, \\\label{j25}
&&c_1\rho\gamma^2[-\frac{\iota\omega}{\alpha}
+u\{a_z+\gamma^2(VV'+uu')\}]
+c_3[-\frac{2\iota\omega}{\alpha}\{(\rho+p)\gamma^4u\nonumber
\end{eqnarray}
\begin{eqnarray}
&&+\frac{B^2\lambda}{4\pi}(u\lambda-V)\}+\iota
k\{(\rho+p)\gamma^4u^2+\frac{\lambda^2
B^2}{4\pi}\}+(\rho+p)\gamma^2\{(a_z\nonumber\\
&&+2\gamma^2uu')
(1+2\gamma^2u^2)+\gamma^2VV'(1+4\gamma^2u^2)\}+\frac{\lambda
B^2}{4\pi}(3\lambda a_z+\lambda')]\nonumber\\
&&+c_4[-\frac{2\iota\omega}{\alpha}\{(\rho+p)\gamma^4V
-\frac{B^2}{4\pi}(u\lambda-V)\}+\iota
k\{(\rho+p)\gamma^4Vu-\frac{\lambda
B^2}{4\pi}\}\nonumber\\
&&+(\rho+p)\gamma^4u\{V'(1+4\gamma^2V^2)+2V(a_z+2\gamma^2uu')\}
-\frac{3B^2}{4\pi}a_z\lambda]\nonumber\\
&&+c_6\frac{B^2}{4\pi}[-\frac{\iota\omega}{\alpha}
\{\lambda+2u(u\lambda-V)\}+\iota ku\lambda+3a_z(2\lambda u-V)
+2u'\lambda\nonumber\\
&&+u\lambda'-V']+c_2p[-\frac{\iota\omega}{\alpha}(\gamma^2-1)+u\gamma^2\{a_z
+\gamma^2(VV'+uu')\}]=0.
\end{eqnarray}
Equations (\ref{j20}) and (\ref{j21}) give $c_5=0$ which shows
that the $z$-component of magnetic field is not affected by
gravity.

\subsection{Numerical Solutions}

We assume the same values of time lapse, $x$ and $z$-components of
the fluid velocity, $\lambda$ and $B$ as given in section
\textbf{6.1}. Using these values with the assumption $\rho=p$, the
mass conservation law in three dimensions gives
$\rho=-\frac{1}{2u}=p$. We obtain a complex dispersion relation
quintic in $k$, i.e., of the form
\begin{eqnarray}\setcounter{equation}{1}\label{v11}
&&A_1(z)k^4+A_2(z,\omega)k^3+A_3(z,\omega)k^2+A_4(z,\omega)k+A_5(z,\omega)
\nonumber\\
&&+\iota\{A_6(z)k^5+A_7(z,\omega)k^4+A_8(z,\omega)k^3
+A_9(z,\omega)k^2+A_{10}(z,\omega)k\nonumber\\
&&+A_{11}(z,\omega)\}=0,
\end{eqnarray}
which can not be solved to get the exact solutions. We have found
the numerical solutions with the help of software
\emph{Mathematica}. We obtain five complex values of $k$ shown in
Figures \textbf{17}-\textbf{21}.

\begin{figure}
\center \epsfig{file=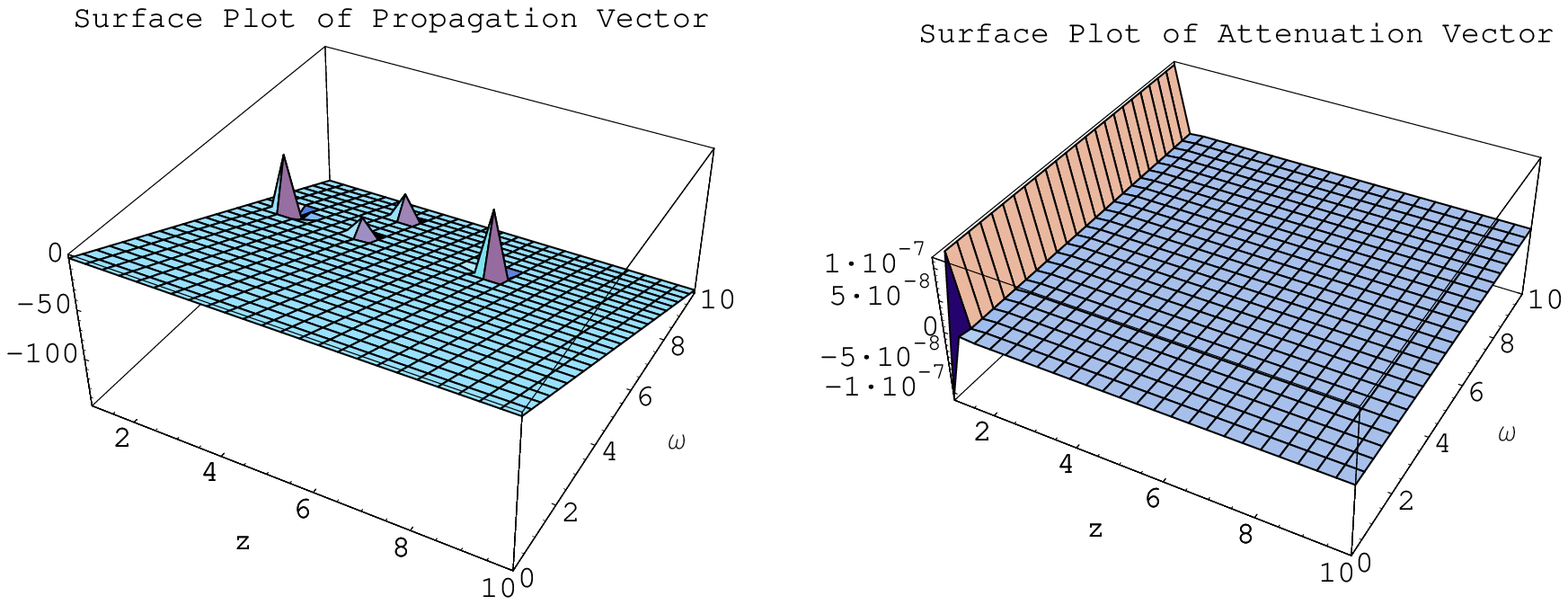,width=0.7\linewidth} \center
\epsfig{file=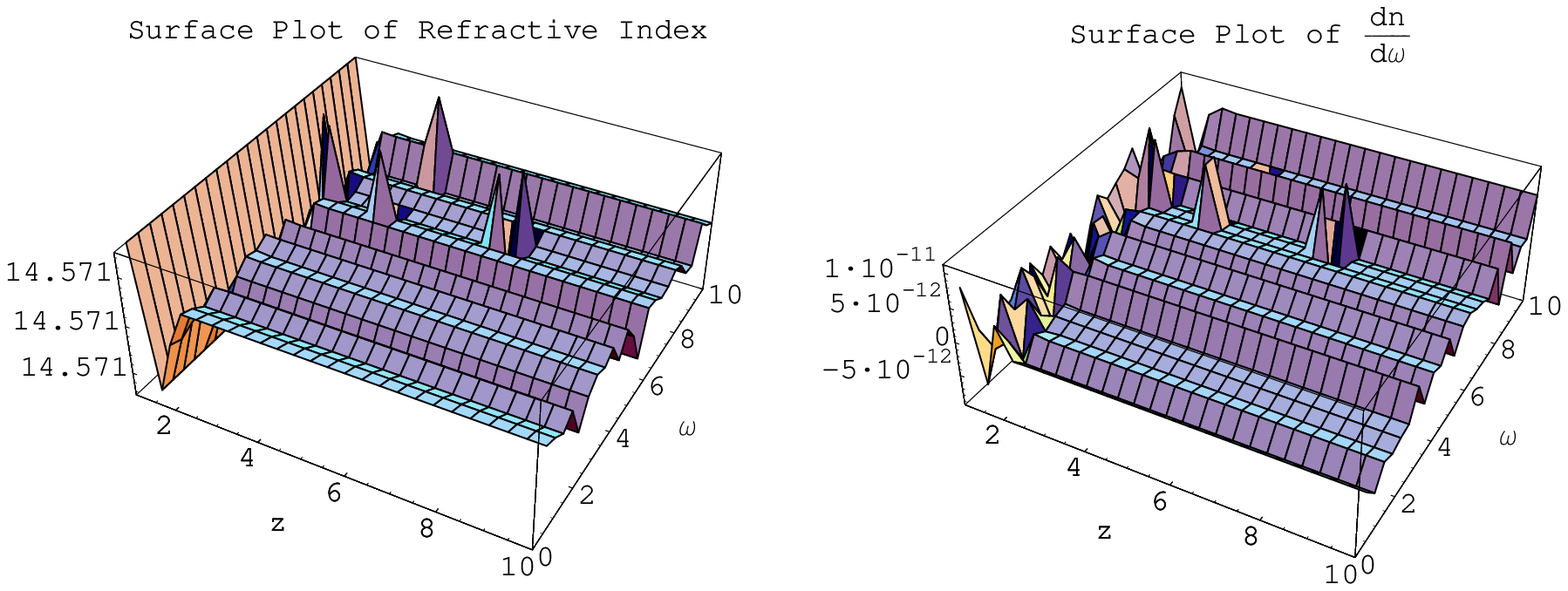,width=0.7\linewidth} \caption{The waves are
directed towards the event horizon. Normal dispersion is found at
random points}
\end{figure}
Figure \textbf{17} shows that the propagation vector decreases
with the increase in angular frequency. In a small region near the
event horizon, the attenuation vector increases as $z$ decreases.
The refractive index is greater than one and its variation with
respect to $\omega$ is positive at random points which shows
normal dispersion of the waves.
\begin{figure}
\center \epsfig{file=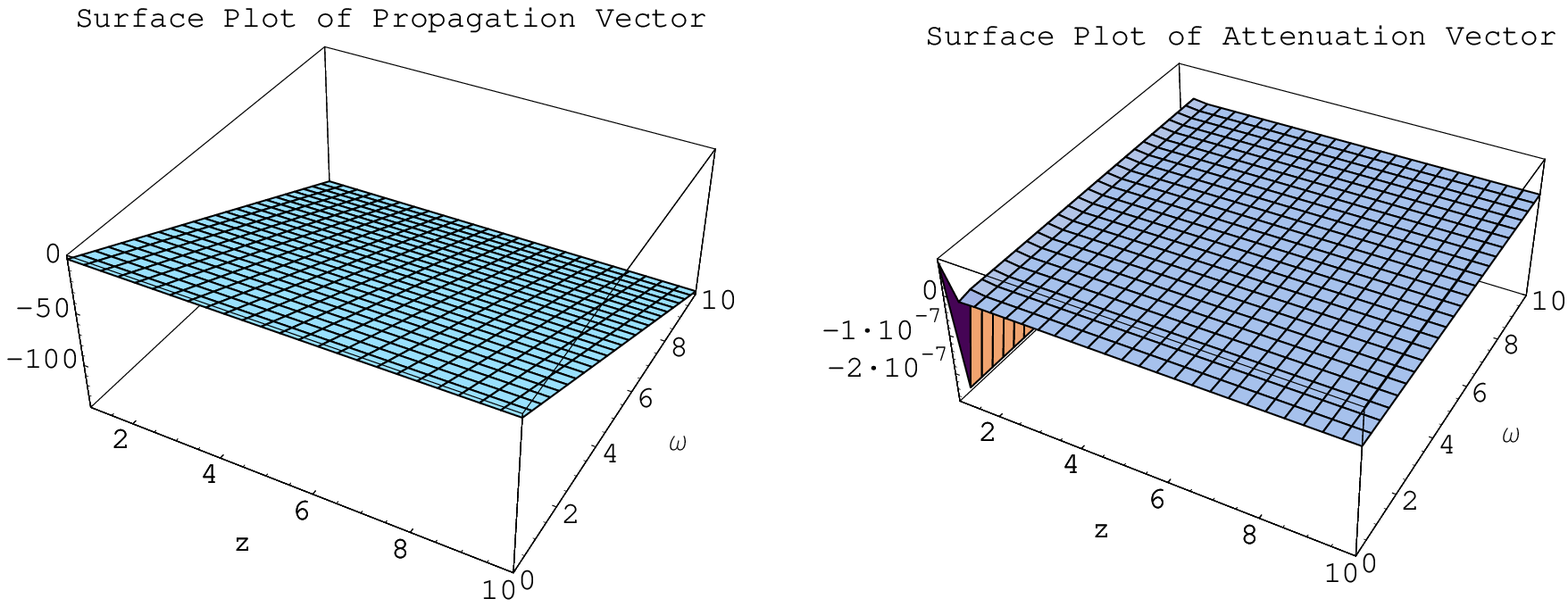,width=0.7\linewidth} \center
\epsfig{file=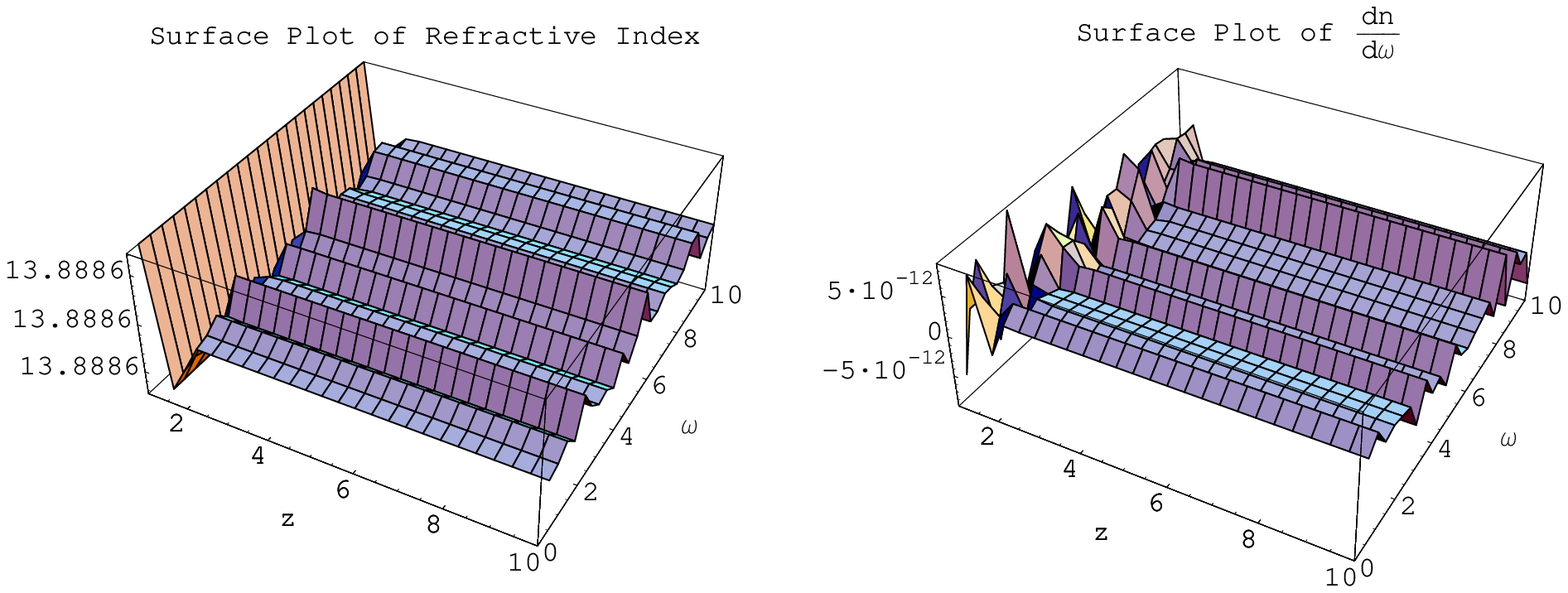,width=0.7\linewidth} \caption{The waves move
towards the event horizon. The waves disperse normally at random
points}
\end{figure}
Figure \textbf{18} shows that the propagation vector decreases
with the increase in angular frequency. The attenuation vector
randomly increases and decreases in the region $1\leq
z\leq2.2,~0\leq\omega\leq10$. The refractive index is greater than
one and its variation with respect to $\omega$ is positive at
random points which shows that the waves disperse normally at
those points.
\begin{figure}
\center \epsfig{file=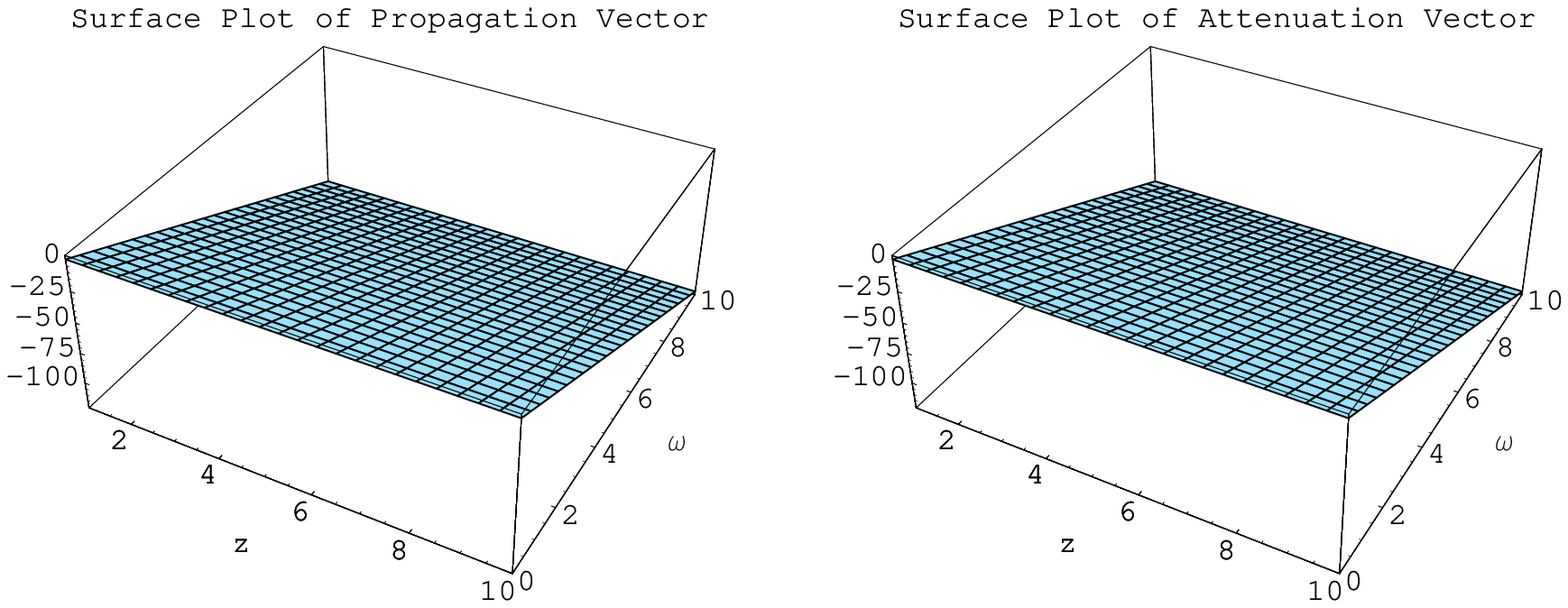,width=0.7\linewidth} \center
\epsfig{file=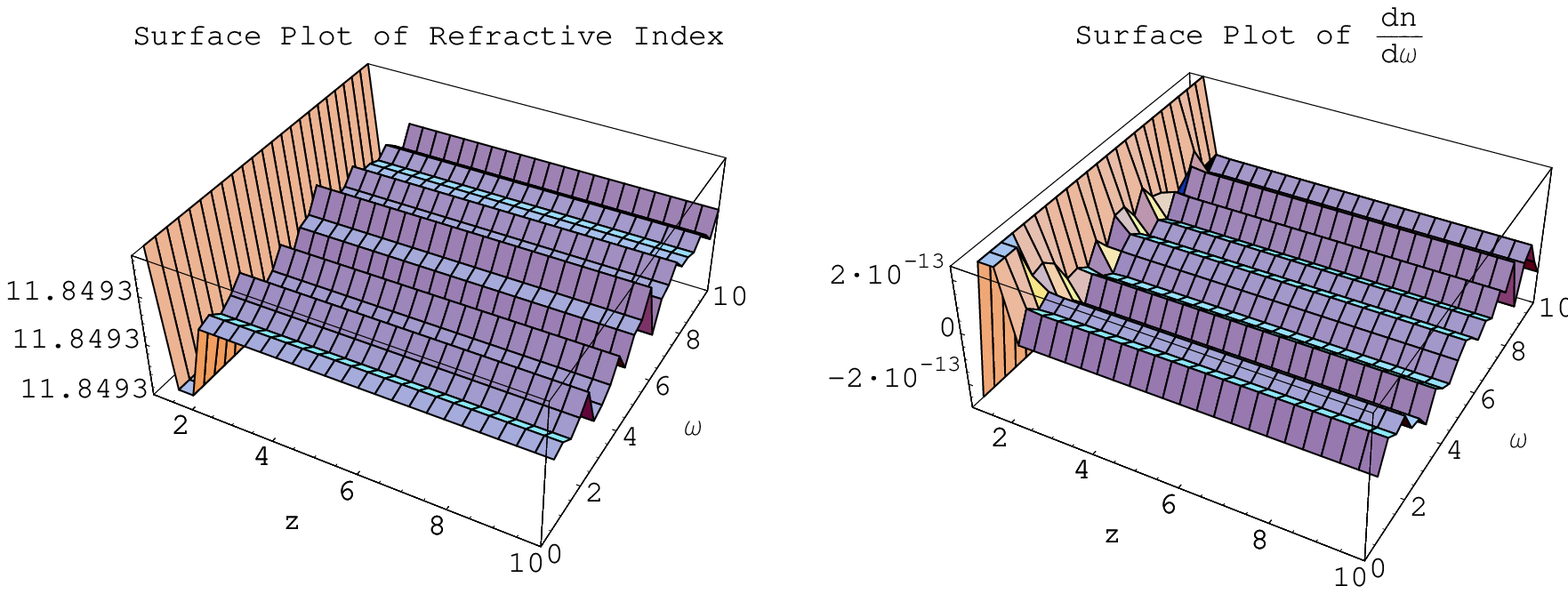,width=0.7\linewidth} \caption{The waves move
towards the event horizon. Dispersion is found to be normal at
random points}
\end{figure}
\begin{figure}
\center \epsfig{file=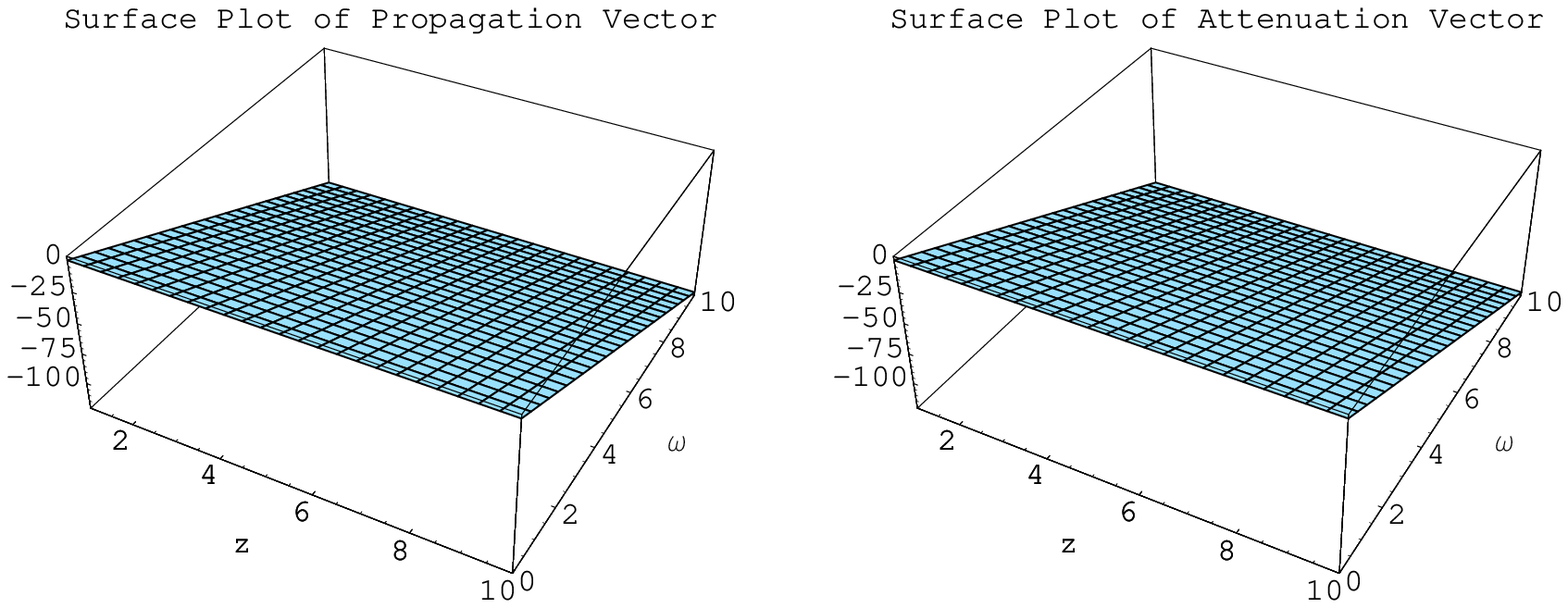,width=0.7\linewidth} \center
\epsfig{file=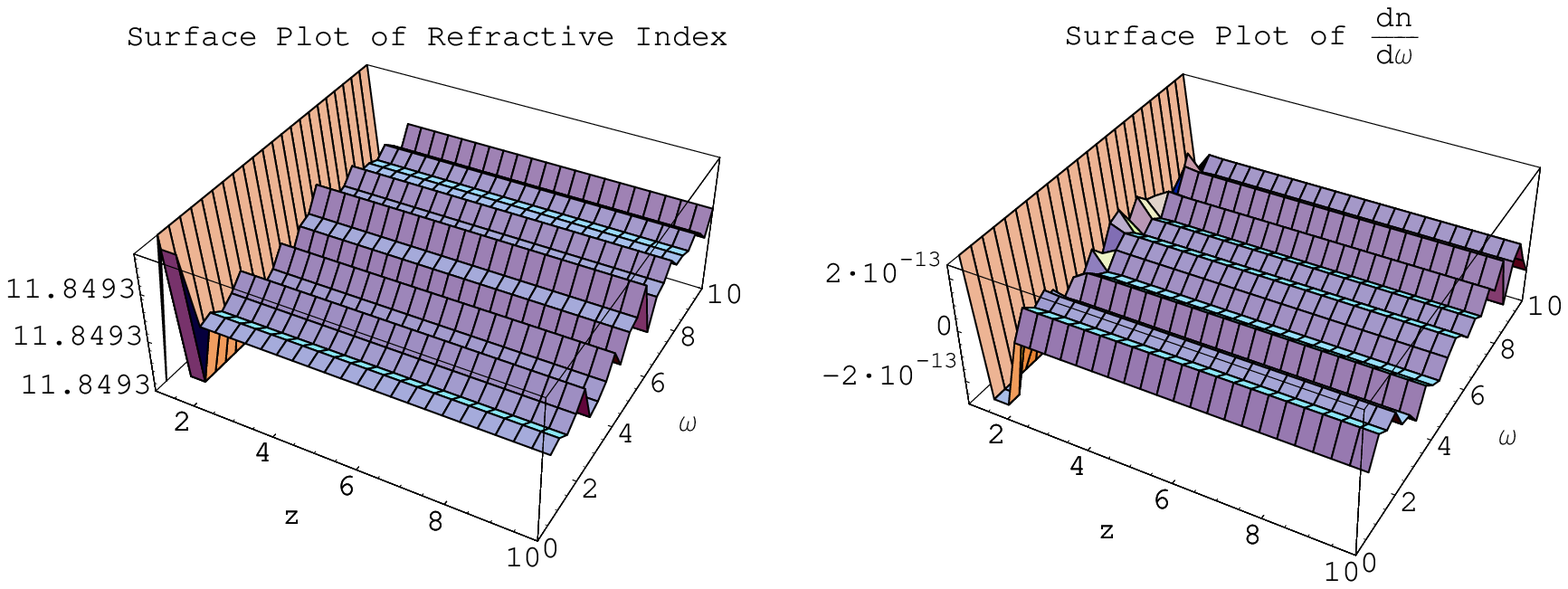,width=0.7\linewidth} \caption{The waves move
towards the event horizon. Normal dispersion is found in a small
region near the event horizon}
\end{figure}
Figure \textbf{19} indicates that both the propagation and
attenuation vectors are negative and decrease with the increase in
angular frequency. The refractive index is greater than one and
its variation with respect to $\omega$ is positive at random
points which shows that the waves disperse normally at those
points. Figure \textbf{20} indicates that both the propagation and
the attenuation vectors decrease with the increase in angular
frequency. The refractive index is greater than one and its
variation with respect to $\omega$ is greater than zero in the
region $1\leq z\leq 1.2,~0\leq \omega\leq 10$ which shows normal
dispersion. Random points of normal and anomalous dispersion are
found otherwise.
\begin{figure}
\center \epsfig{file=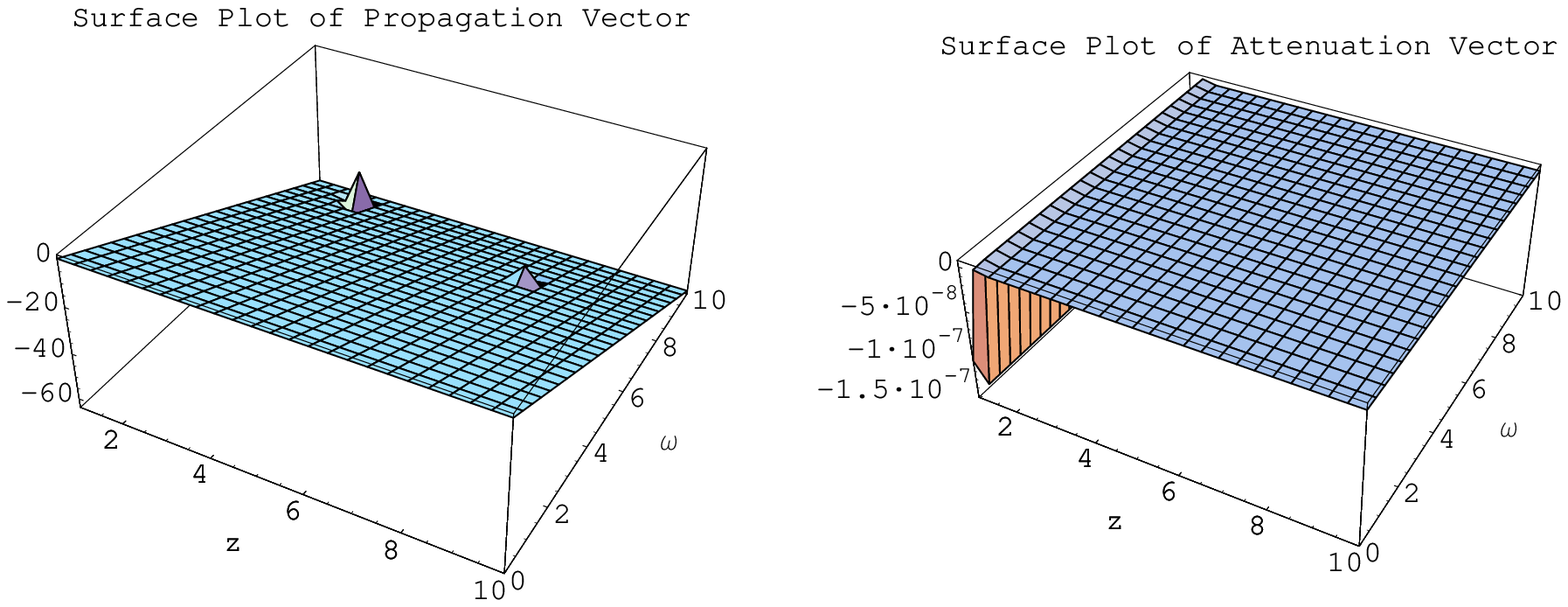,width=0.7\linewidth} \center
\epsfig{file=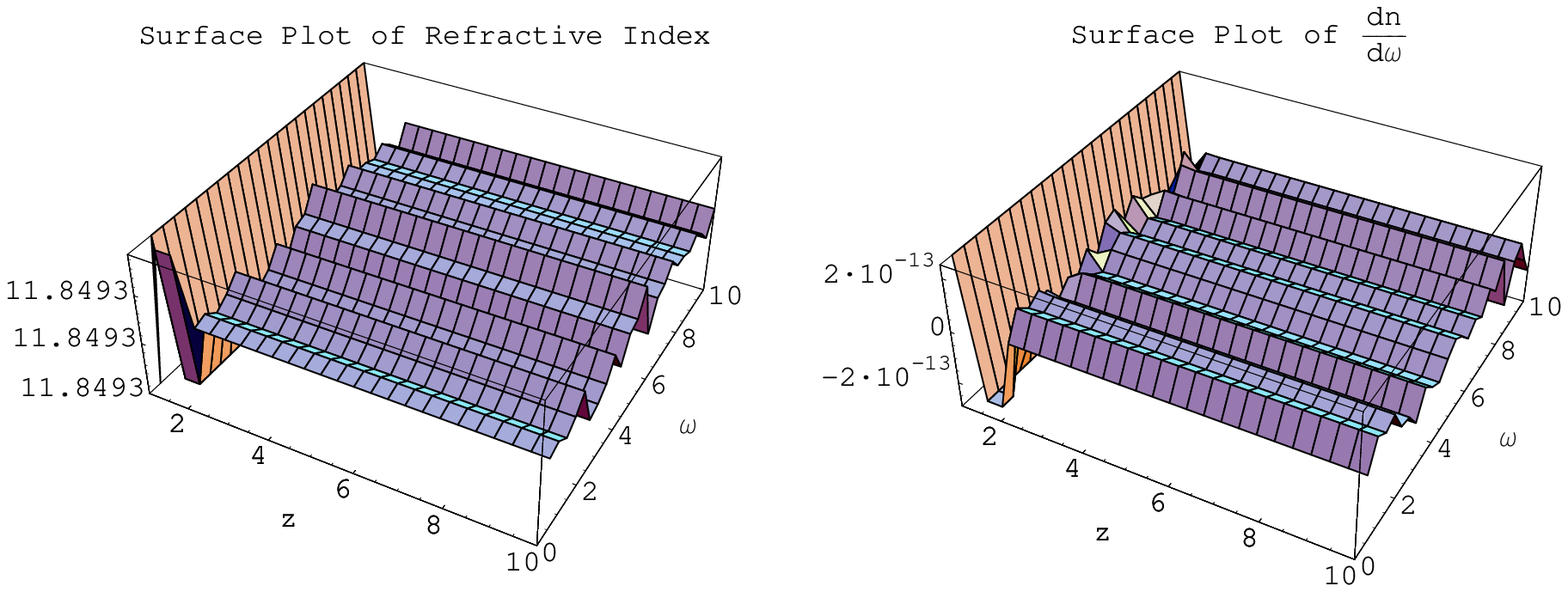,width=0.7\linewidth} \caption{The waves are
directed towards the event horizon. Normal dispersion is found at
random points }
\end{figure}
Figure \textbf{21} shows that the propagation vector decreases
with the increase in angular frequency. The attenuation vector
increases and decreases randomly in the region $1\leq z\leq
2.2,~0\leq \omega\leq 10$. The refractive index is greater than
one and its variation with respect to $\omega$ is positive near
the event horizon which indicates normal dispersion of waves.

In Figures \textbf{17}-\textbf{21}, the propagation vector admits
negative values indicating the direction of waves towards the
event horizon. Furthermore, in all these figures, a small region
near the event horizon admits the increase in refractive index
with the decrease in $z$.

\section{Outlook}

We have discussed the cold and isothermal plasma wave properties
of the Schwarzschild black hole magnetosphere in 3+1 formalism.

For this purpose, we have used the planar analogue (given by
Eq.(\ref{R})) due to the following reasons:
\begin{itemize}
\item It is difficult to study the MHD stream equation and the
behavior of a perturbed magnetosphere in the Schwarzschild
spacetime. We have chosen a different way and altered the
spacetime (i.e., we have taken the planar analogue) which
simplifies the analysis of magnetosphere. It has been done while
preserving the following key features of the Schwarzschild metric:
\begin{enumerate}
\item The behavior of the Schwarzschild time coordinate as $r\rightarrow
2M$ is analogous to the behavior of Schwarzschild planar analogue
time coordinate as $z\rightarrow 0$.

\item The planar spacetime and its stationary MHD magnetosphere
have nice computational features that all aspects of the
magnetosphere should become asymptotically $z$ independent at
large $z$.
\end{enumerate}

\item The planar analogue is not empty which is not important for
our analysis because it serves merely as a testbed for studying
various aspects of the interactions of gravity with plasma
\cite{26}. Therefore, we are free to assume that there is no
direct non-gravitational interaction between the plasma and
materials whose stress-energy produce the spacetime curvature.
\end{itemize} The plasma is assumed to be living in this
planar analogue for which the dispersion relations are obtained
from the Fourier analyzed perturbed 3+1 GRMHD equations for the
non-rotating, rotating non-magnetized and rotating magnetized
backgrounds. These relations give the value of $z$-component of
the wave vector which yields the propagation and attenuation
vectors, the refractive index and its change with respect to
angular frequency. The graphs are obtained to have a clear
viewpoint. A summary of
the results is given below.

For the \textbf{cold plasma} living in the vicinity of Schwarzschild
regime, normal dispersion of waves is found in the Figures
\textbf{1}, \textbf{3} and \textbf{8} which indicate that the waves
can pass through the medium.  Figures \textbf{2}, \textbf{4},
\textbf{6} and \textbf{9} represent that the waves disperse
anomalously in most of the region. Random dispersion of waves is
found in the Figures \textbf{5} and \textbf{7}. Figure \textbf{3}
shows that the waves damp in a small region near the event horizon.
In Figures \textbf{4} and \textbf{5}, the waves grow near the event
horizon in a small region. In Figure \textbf{9}, the attenuation
vector increases with the decrease in $z$ in a small region near the
event horizon. In Figures \textbf{1}-\textbf{9}, the waves damp and
grow randomly with the increase in angular frequency.

For the \textbf{isothermal plasma} surrounding the Schwarzschild
black hole, Figure \textbf{14} admits normal dispersion near the
event horizon. The waves disperse anomalously in Figure
\textbf{11}. In Figure \textbf{10}, the attenuation vector
increases whereas it decreases for Figure \textbf{11} with the
increase in angular frequency. It shows that the waves damp in
Figure \textbf{10} and grow for Figure \textbf{11} with the
increase in angular frequency. Figure \textbf{12} shows that the
wave grow and damp randomly with the increase in angular
frequency. In Figures \textbf{15} and \textbf{16}, the waves grow
in a small region near the event horizon. In Figure \textbf{17},
the waves damp whereas they grow in Figures \textbf{18} and
\textbf{21} with the decrease in $z$ in small regions near the
event horizon. In Figures \textbf{19} and \textbf{20}, the
attenuation vector decreases with the increase in angular
frequency.

For the cold plasma living in the non-rotating Schwarzschild
background, there exists a case where the waves disperse normally
(Figure \textbf{1}) whereas for the isothermal plasma, normal
dispersion lies at random points. In the rotating non-magnetized
background, the cold plasma shows normal dispersion in Figure
\textbf{3} whereas the isothermal plasma admits normal dispersion
in a small region near the event horizon in Figure \textbf{14}. In
the rotating magnetized background, Figure \textbf{8} indicates
that the cold plasma admits normal dispersion in most of the
region whereas the isothermal plasma admits normal dispersion at
random points. Thus we can conclude that the pressure ceases the
normal dispersion of waves.

In most of the cases, the refractive index increases as the waves
move towards the event horizon in a small region near the event
horizon. This shows that the refraction of waves increases as they
move towards the event horizon. For all the backgrounds of cold and
isothermal plasma, the propagation vector takes negative values
which shows that the waves move towards the event horizon.\\\\
{\bf Acknowledgment:} We would like to thank Miss Umber Skeikh for
the fruitful discussions during this work.

\end{document}